\documentclass[twocolumn,twocolappendix]{aastex631}
\usepackage{placeins}
\usepackage{amsmath} 
\usepackage{multirow}
\usepackage{subfigure}
\usepackage{array} % For p{width} columns

\begin{document}

%\title{Coronal Heating Driven by Oscillatory Reconnection}
% \title{Caracterisation of the oscillaor}
%\title{Bridging 2D and 3D oscillatory reconnection}

%\title{The periodicity of three-dimensional oscillatory reconnection}

% \title{Three-dimensional oscillatory reconnection: periodicity and self-similarity}
\title{The periodicity of three-dimensional oscillatory reconnection}

\author[0000-0002-5082-1398]{Luiz A. C. A. Schiavo}
\affiliation{Northumbria University, Newcastle upon Tyne, NE1 8ST, UK}
%\affiliation{Department of Mathematics, Physics \& Electrical Engineering, Northumbria University, Newcastle upon Tyne, NE1 8ST, UK}

\author[0000-0002-5915-697X]{Gert J. J. Botha}
\affiliation{Northumbria University, Newcastle upon Tyne, NE1 8ST, UK}
%\affiliation{Department of Mathematics, Physics \& Electrical Engineering, Northumbria University, Newcastle upon Tyne, NE1 8ST, UK}

\author[0000-0002-7863-624X]{James A. McLaughlin}
%\affiliation{Department of Mathematics, Physics \& Electrical Engineering, Northumbria University, Newcastle upon Tyne, NE1 8ST, UK}
\affiliation{Northumbria University, Newcastle upon Tyne, NE1 8ST, UK}

%% Note that the \and command from previous versions of AASTeX is now
%% depreciated in this version as it is no longer necessary. AASTeX 
%% automatically takes care of all commas and "and"s between authors names.

%% AASTeX 6.31 has the new \collaboration and \nocollaboration commands to
%% provide the collaboration status of a group of authors. These commands 
%% can be used either before or after the list of corresponding authors. The
%% argument for \collaboration is the collaboration identifier. Authors are
%% encouraged to surround collaboration identifiers with ()s. The 
%% \nocollaboration command takes no argument and exists to indicate that
%% the nearby authors are not part of surrounding collaborations.

%% Mark off the abstract in the ``abstract'' environment. 
\begin{abstract}
%context
%aim
%method
%results
%conclusion
%context
{
Oscillatory reconnection is a dynamic, magnetic relaxation mechanism in which a perturbed null point reverts back to equilibrium via time-dependent reconnection. In this paper, we investigate the long-term periodic signal generated by a three-dimensional (3D) magnetic null point, when it is perturbed by a non-periodic driver, for a variety of driving amplitudes. We solve the 3D nonlinear magnetohydrodynamic (MHD) equations using a bespoke numerical boundary condition (a sponge region) that damps wave reflections and thus allows the long-term periodic signal at the 3D null point to be investigated. We observe multiple cycles of the 3D oscillatory reconnection mechanism for the first time. We find that the periodicity is both constant and independent of the choice of driving amplitude. Furthermore, the resultant time-dependent current density at the null point normalized by the driving amplitude is invariant. We extract a single period for oscillatory reconnection at a 3D null point, opening the future possibility of using this characteristic period as a diagnostic tool to reveal indirectly the fundamental plasma properties of 3D null points. 
}

\end{abstract}

%% Keywords should appear after the \end{abstract} command. 
%% The AAS Journals now uses Unified Astronomy Thesaurus concepts:
%% https://astrothesaurus.org
%% You will be asked to selected these concepts during the submission process
%% but this old "keyword" functionality is maintained in case authors want
%% to include these concepts in their preprints.
\keywords{Solar magnetic reconnection(1504) --- Solar physics(1476) --- Solar coronal transients(312) --- Solar coronal heating(1989) --- Magnetohydrodynamics(1964)}

%% From the front matter, we move on to the body of the paper.
%% Sections are demarcated by \section and \subsection, respectively.
%% Observe the use of the LaTeX \label
%% command after the \subsection to give a symbolic KEY to the
%% subsection for cross-referencing in a \ref command.
%% You can use LaTeX's \ref and \label commands to keep track of
%% cross-references to sections, equations, tables, and figures.
%% That way, if you change the order of any elements, LaTeX will
%% automatically renumber them.
%%
%% We recommend that authors also use the natbib \citep
%% and \citet commands to identify citations.  The citations are
%% tied to the reference list via symbolic KEYs. The KEY corresponds
%% to the KEY in the \bibitem in the reference list below. 

\section{Introduction} \label{sec:intro}
Magnetic reconnection serves as a critical energy conversion mechanism in plasma systems, efficiently transforming stored magnetic energy into thermal and kinetic energy while facilitating particle acceleration and topological magnetic field restructuring \citep[e.g.][]{2022LRSP...19....1P,BROWNING2024100049}. This fundamental process drives key solar phenomena, particularly in generating coronal mass ejections \citep[e.g.][]{2012LRSP....9....3W,2024ApJ...975..168W}, the energy release during solar flares \citep[e.g.][]{2017LRSP...14....2B,2025SSRv..221...27D} and chromospheric anemone jet observations demonstrate how small-scale reconnection events in the lower solar atmosphere may contribute to chromospheric and coronal heating processes \citep[e.g.][]{Shibata2007,2024ApJ...962L..35S}. An overview of the outstanding challenges in understanding magnetic reconnection can be found in \citet{2023BAAS...55c.192J}, \citet{2024mpsp.book..345P} and \citet{2025SSRv..221...17N}.

Reconnection theory has developed along four main research directions: (i) kinetic-scale collisionless effects  \cite[e.g.][]{2025SSRv..221...20G}, (ii) the extension of established two-dimensional (2D) models to three-dimensional (3D) configurations \citep{priest_three-dimensional_2009,pontin_magnetic_2022}, (iii) transient and time-dependent behavior \citep{Thurgood2017,2025SSRv..221...16L}, and (iv) the dynamics of local-global system interactions \citep[e.g.][]{2024arXiv240605901S}. Our investigation will concentrate on topics (ii) and (iii), utilizing resistive magnetohydrodynamic (MHD) simulations to enhance understanding in these areas.

Oscillatory Reconnection (OR) represents a distinct class of time-dependent magnetic reconnection that exhibits periodic variations in magnetic connectivity. This phenomenon was first observed by \cite{Craig1991} through their investigation of magnetic field relaxation in 2D  X-point configurations. What makes OR particularly noteworthy is its self-sustaining periodicity -- the oscillatory behavior emerges inherently from the system's relaxation dynamics rather than requiring periodic external driving. This characteristic allows OR to produce regular, periodic outputs even when initiated by aperiodic perturbations \citep{McLaughlin2009}.

OR has garnered significant interest as a potential mechanism driving quasi-periodic pulsations (QPPs) in the impulsive and decaying phase of solar flares \citep[e.g.][]{hayes_quasi-periodic_2016,2024A&A...684A.215C} and stellar flares, \citep[e.g.][]{2018MNRAS.475.2842D}. QPPs, characterized by oscillatory or pulsating signatures in flare emission, are frequently observed across multiple wavelengths, including microwave emissions \citep{nakariakov_quasi-periodic_2018}, extreme ultraviolet  \citep[e.g.][]{dominique_detection_2018,2025JGRA..13033772L}, soft and hard X-rays \citep[e.g.][]{dennis_detection_2017,shi_multiwavelength_2024} and gamma-ray \citep{nakariakov_quasi-periodic_2010}. QPPs typically exhibit periods ranging from seconds to minutes and amplitudes of approximately 1–10\% of signal amplitude. Evidence in some events suggests that QPPs may be generated by oscillatory energy injection into the reconnection region \citep{yuan_compact_2019} at a loop top, supporting the role of time-dependent magnetic reconnection. Despite the growing number of observational studies, the physical mechanisms responsible for QPP generation remain unresolved, with several competing models proposed \citep[see reviews by][]{McLaughlin2018,Zimovets2021}. OR has emerged as a promising candidate, offering a plausible explanation for the periodic modulation of flare energy release. OR has also been reported in flux rope formation observations, e.g. \citet{2019ApJ...874L..27X}, in breakout reconnection preceding a solar jet \citep{hong_observation_2019}, in a formation of an intermediate filament \citep{sun_formation_2023} and at a coronal bright point \citep{2025arXiv250513859H}.

In 2D and 2.5D, OR has been observed in MHD simulations across a wide variety of systems, such as in the 2D X-point configuration \citep{McLaughlin2009,
%2020NatSR..1015603S,
Karampelas2023,Talbot2024,schiavo2024energymap}, and arcade configurations \citep{Tarr2017,Santamaria2018}, as well as during the coalescence of magnetic flux ropes \citep{Stewart2022,schiavo2024PoP}, and in the emergence of a magnetic flux tube from the convection zone \citep{wang_numerical_2025}.

In 3D, magnetic reconnection can occur in current layers either at 3D null points or in their absence. In either scenario, the evolution of field lines is characterized by continuous slippage rather than a one-to-one cut-and-paste of field line pairs \citep{priest_nature_2003}. Reconnection at 3D null points can occur in various modes, including spine-fan reconnection,  torsional spine reconnection, and torsional fan reconnection \citep{priest_three-dimensional_2009}. Studies on 3D magnetic reconnection have focused primarily on analytical models \citep{priest_three-dimensional_2009}, steady-state models, \citep{wyper_torsional_2010,wyper_torsional_2011}, or simulations of the tearing instability \citep{wyper_dynamic_2014,huang_turbulent_2016}. 

\cite{Thurgood2017} pioneered the study of 3D OR, demonstrating that reconnection at a fully 3D null point can occur in a natural, time-dependent, and periodic manner. They examined a 3D null point configuration and disturbed the system with a spherical implosion that triggered OR. Their work revealed the reorientation of the current sheet at the null point for a single oscillation period. However, due to an absence of any numerical procedure to handle reflected waves at the simulation boundaries, their investigation stopped after a single OR cycle. \cite{sabri_plasma_2021,sabri_propagation_2022} also simulate a similar configuration to \cite{Thurgood2017} using an Alfvén wave as a driver, but again the simulation was constrained to a short simulation time due to the interference of the boundaries.

In 2D, it has been shown that the oscillations of the current sheet occur periodically and decrease in amplitude \citep{Karampelas2023,Talbot2024,schiavo2024energymap} and are independent of the initial pulse \citep{Karampelas2022b}. It is unclear if these 2D properties carry over to 3D, such as whether or not the period in 3D OR is constant over time, as its 2D counterpart, and how the current density oscillation decays over time.

This study aims is analyze and quantify the OR phenomenon in 3D, by investigating the long-term current density oscillation at the null point. We will build upon the pioneering work of \cite{Thurgood2017} to analyze multiple periods OR and the corresponding amplitude decay. We will also investigate the sensitivity of the system to the initial pulse strength to determine if OR in 3D remains independent of the initial driver, similar to its 2D counterpart \citep{Karampelas2022b}.

The paper is organized as follows: \S\ref{sec:numerical-model} details our numerical approach; \S\ref{sec:results} details our findings: 
the transient evolution (\S\ref{sec:evolution}),  pulse sensitivity (\S\ref{sec:evolution-drivers}), the characterization of the OR signal (\S\ref{sec:jy-at-null}), OR modeling (\S\ref{sec:jy-modeling}) and an analysis of vorticity evolution(\S\ref{sec:circulation}); with conclusions given in \S\ref{sec:conclusions}.

%%%%%%%%%%%%%%%%%%%%%%%%%%%%%%%%%%%%%%%%%%%%%%%%%%%%%%%%%%
%%%%%%%%%%%%%%%%%%%%%%%%%%%%%%%%%%%%%%%%%%%%%%%%%%%%%%%%%%
%%%%%%%%%%%%%%%%%%%%%%%%%%%%%%%%%%%%%%%%%%%%%%%%%%%%%%%%%%

\section{Numerical model} \label{sec:numerical-model}
\subsection{Governing equations}\label{sec2.1}
We solve the 3D resistive MHD equations through the utilization of the Lare3D code \citep{Arber2001}. The equations are solved in Lagrangian form, employing a Lagrangian-Eulerian remap procedure and can be expressed in dimensionless form as follows:
\begin{eqnarray}
\frac{D\rho}{D t} &=& - \rho \nabla \cdot \mathbf{v} , \label{eq:mass}\\
\frac{D\mathbf{v}}{D t} &=& \frac{1}{\rho}(\nabla \times \mathbf{B} ) \times \mathbf{B} - \frac{1}{\rho}\nabla p  +\frac{1}{\rho} \mathbf{F}_{visc}, \label{eq:momentuum}\\
\frac{De}{D t} &=& - \frac{p}{\rho} \nabla \cdot \mathbf{v} + \frac{\eta}{\rho}|\mathbf{j}|^2+ \frac{1}{\rho} Q_{visc}, \label{eq:energy}\\
\frac{D\mathbf{B}}{D t} &=& (\mathbf{B}\cdot\nabla) \mathbf{v} - \mathbf{B}(\nabla \cdot \mathbf{v}) - \nabla\times (\eta\nabla\times \mathbf{B}) , \label{eq:induction}\\
p &=& \rho e(\gamma -1) ,
\label{eq:mhd}
\end{eqnarray}
%\noindent 
where $D/Dt$ represents the material derivative, $\mathbf{v}$ denotes the velocity vector, $\mathbf{B}$ represents the magnetic field, $\mathbf{j}$ is the current density, $\rho$ signifies plasma density, $p$ corresponds to plasma thermal pressure, $e$ represents specific internal energy, $\eta$ characterizes the resistivity, which is considered uniform, and $\gamma$ is the ratio of specific heats, set to 5/3. To accurately accommodate steep gradients such as shocks and address numerical instabilities, Lare3D utilizes a numerical viscosity \citep{Arber2001} which is implemented by adding a forcing term $\mathbf{F}_{visc}$ in the momentum equation  and its corresponding heat, $Q_{visc}$, in the energy equation. For accurate shock-capturing, we set the numerical viscosity parameters $\nu_1=$ 0.1 and $\nu_1=$ 0.5 in our Lare3D model, where more details about the shock capturing scheme and its calibration can be found in \citet{Caramana1998}.

The model assumes full ionization of the plasma and non-dimensionalizes the governing equations with respect to length scale $L_0$, magnetic field $B_0$, and density $\rho_0$. These constants define non-dimensionalization for velocity $v_0 = B_0/\sqrt{\left.\mu_0 \rho_0\right.}$, thermal pressure $p_0=B_0^2/\mu_0$, time $t_0=L_0/v_0$, current density $j_0=B_0/\mu_0 L_0$, specific internal energy $e_0=v_0^2$, temperature $T_0=e_0\overline{m}/k_B$ and resistivity $\eta_0= \mu_0L_0v_0$, where $\mu_0$ is the vacuum magnetic permeability, $k_B$ is the Boltzmann constant and $\overline{m}$ the average mass of ions. 
We set the resistivity as $\eta=10^{-3}\eta_0$. Finally, here the subscript $0$ refers to the non-dimensionalization scales used in Lare3D.

%%%%%%%%%%%%%%%%%%%%%%%%%%%%%%%%%%%%%%%%%%%%%%%%%%%%%%%%%%%%%%%%%%%%%%%%%%

\subsection{Equilibrium magnetic field and initial perturbation}\label{sec:2.2}
The magnetic field configuration consists of a three-dimensional null at the origin of the Cartesian domain. This is known as a linear, proper, potential null point \citep{1996PhPl....3..759P} with the magnetic null point itself located at the origin and where the fan is aligned with the $z=0$, $xy-$plane. In contrast, a spine is primarily aligned with the $z-$axis. Our investigation builds upon the work of \citet{Thurgood2017}. In that paper and here, a 3D X-point is considered that is in equilibrium along with a perturbation field:
\begin{equation}
\mathbf{B}=\overline{\mathbf{B}} + \mathbf{B}^\prime , \label{equation=X-point}
\end{equation}
where the initial state, $\overline{\mathbf{B}}$, and its perturbation, $\mathbf{B}^\prime$, are given by:
\begin{equation}
\overline{\mathbf{B}}=(x,y,-2z), \hspace{0.5cm} \mathbf{B}^\prime = \nabla \times \mathbf{A}^\prime, 
\label{eq:perturbation}
\end{equation}
\begin{equation}
\mathbf{A}^\prime = \psi \exp \left( -\frac{x^2 + y^2 + z^2}{2 \sigma^2} \right) \hat{\bf{y}} .
\label{eq:potential}
\end{equation}
The notation used here is that $\overline{(\ \ )}$ means an initial state and $(\  \ )^\prime$ a perturbation to the initial state. A uniform equilibrium state is used with a density of $\overline{\rho} = 1$, a velocity of $\overline{\mathbf{v}} = 0$, and a pressure of $\overline{p} = 0.005$. This $\overline{p}$ is chosen so that the plasma-$\beta=0.01$ at a distance of unity from the null point. Additionally, the magnetic Reynolds number was set to $R_m = 10^3$. The simulations were conducted over a period of 60 time units (i.e. $60\: t_0$) in every case. 

\begin{figure*}[p]
	\centering
 \begin{subfigure}[$\psi=0.01$]
 {\includegraphics[trim = 30 10 70 10, clip, width=0.32\textwidth]{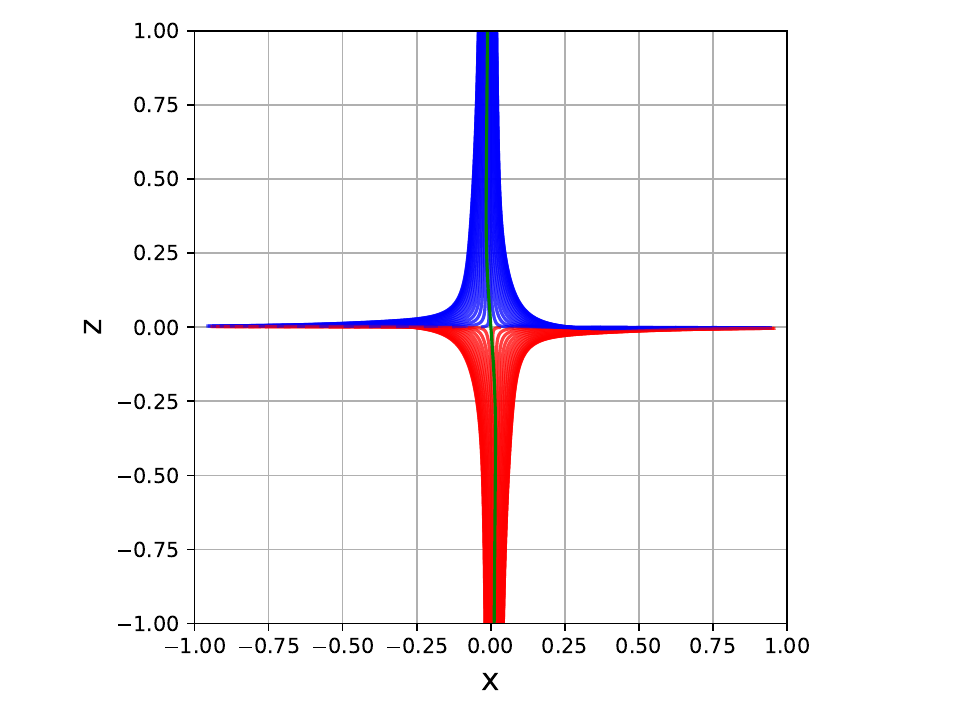}}
 \end{subfigure}
\centering
	 \begin{subfigure}[$\psi=0.01$]
 {\includegraphics[trim = 30 10 70 10, clip, width=0.32\textwidth]{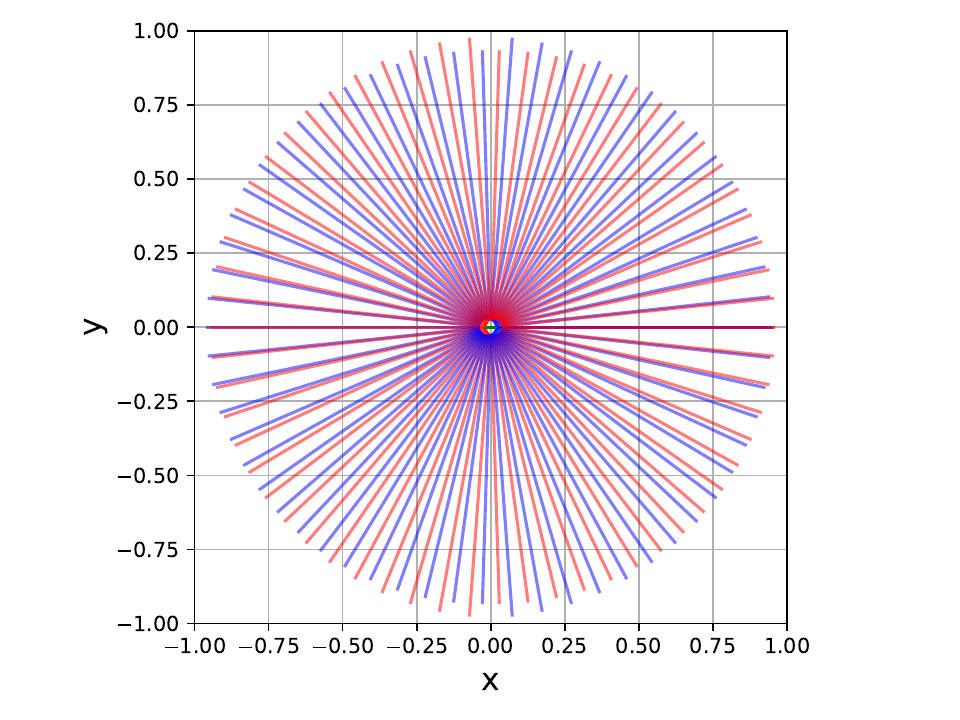}}
 \end{subfigure}
\centering
	\centering
 \begin{subfigure}[$\psi=0.01$]
 {\includegraphics[trim =  70 0 70 30, clip, width=0.32\textwidth]{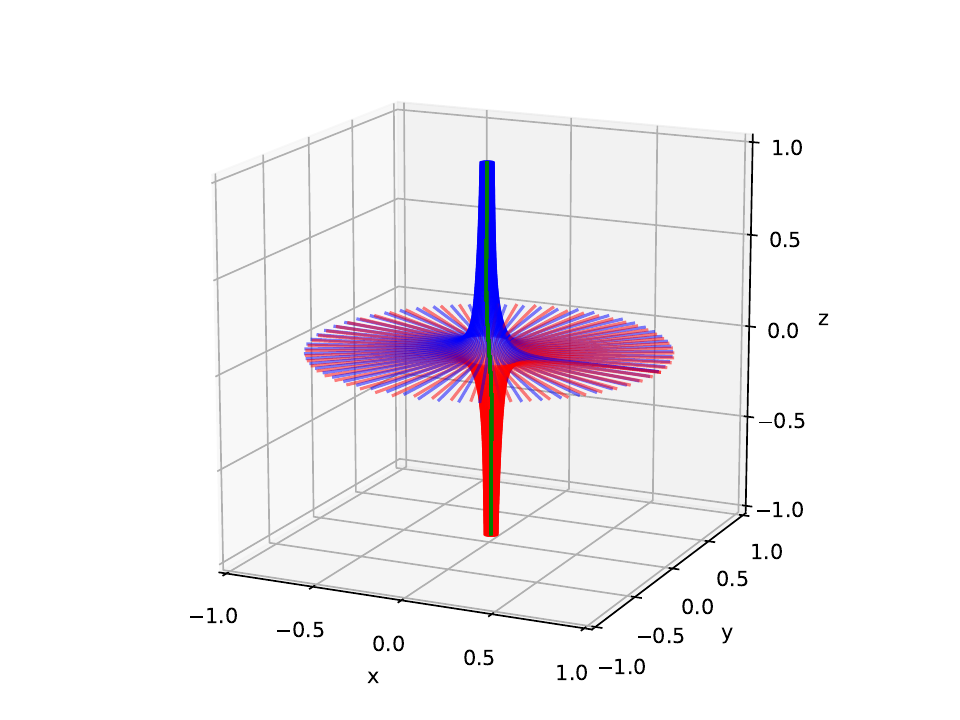}}
 \end{subfigure}

    \centering
 \begin{subfigure}[$\psi=0.05$]
 {\includegraphics[trim = 30 10 70 10, clip, width=0.32\textwidth]{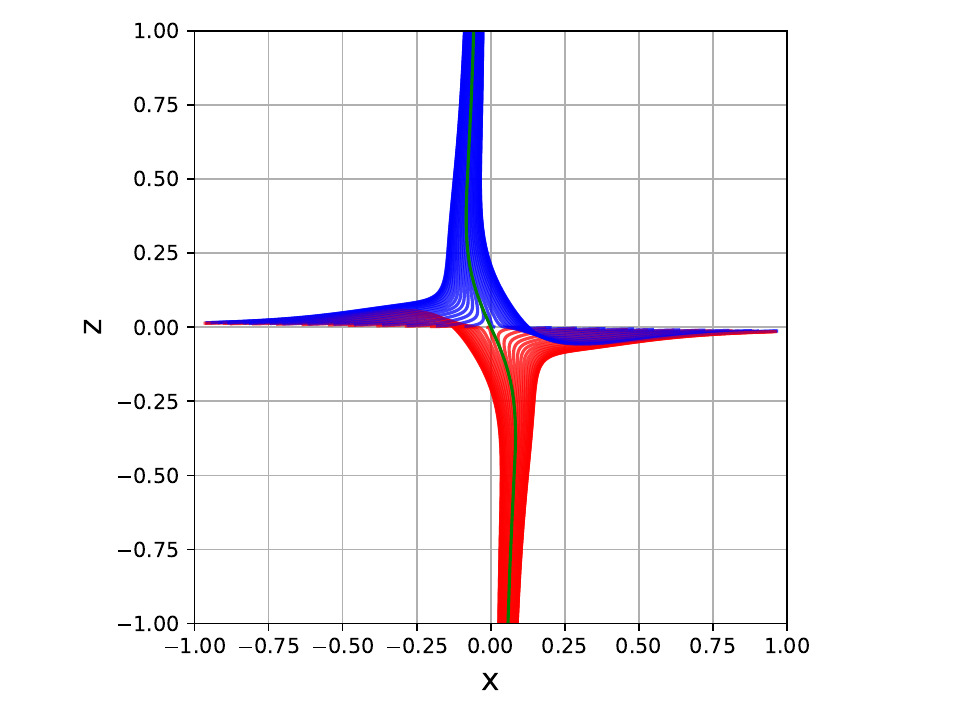}}
 \end{subfigure}
\centering
	 \begin{subfigure}[$\psi=0.05$]
 {\includegraphics[trim = 30 10 70 10, clip, width=0.32\textwidth]{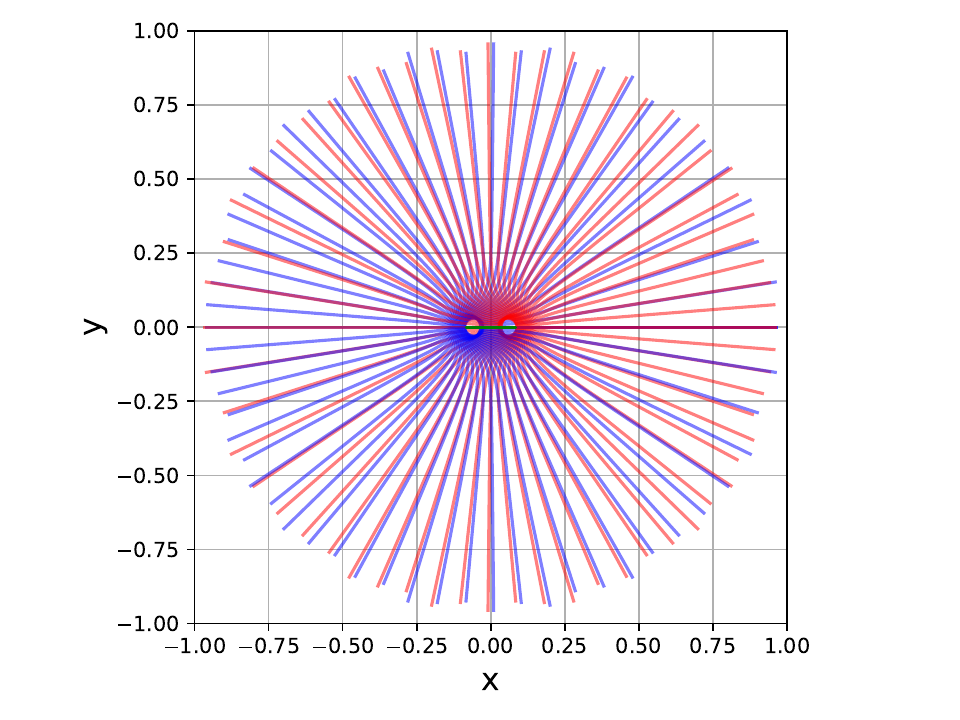}}
 \end{subfigure}
\centering
	\centering
 \begin{subfigure}[$\psi=0.05$]
 {\includegraphics[trim =  70 0 75 30, clip, width=0.32\textwidth]{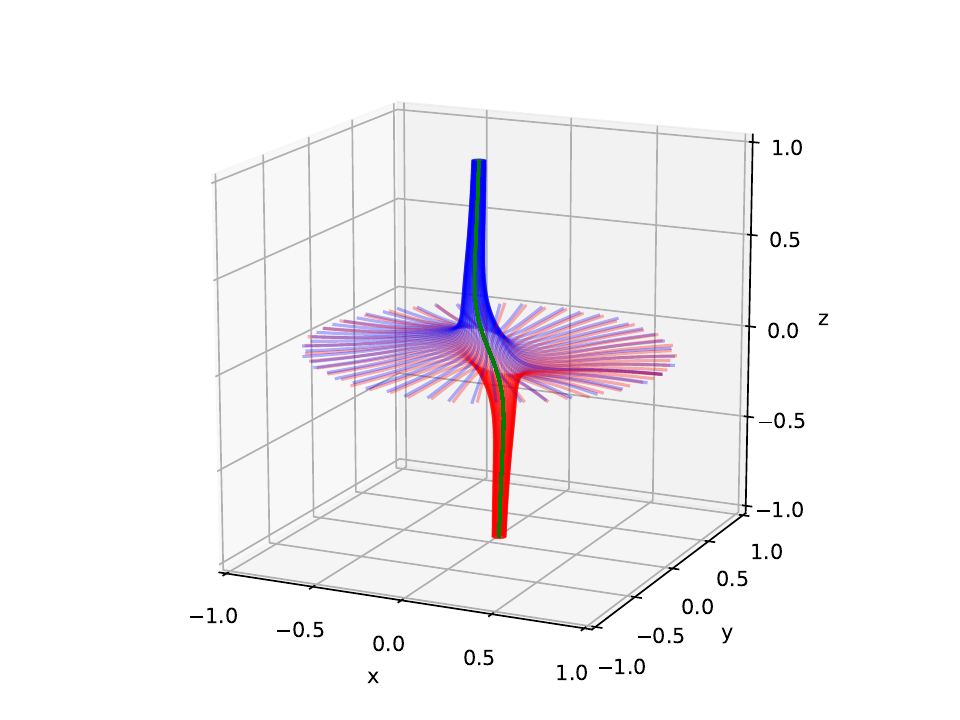}}
 \end{subfigure}
\centering

\centering
	\centering
 \begin{subfigure}[$\psi=0.1$]
 {\includegraphics[trim = 30 10 70 10, clip, width=0.32\textwidth]{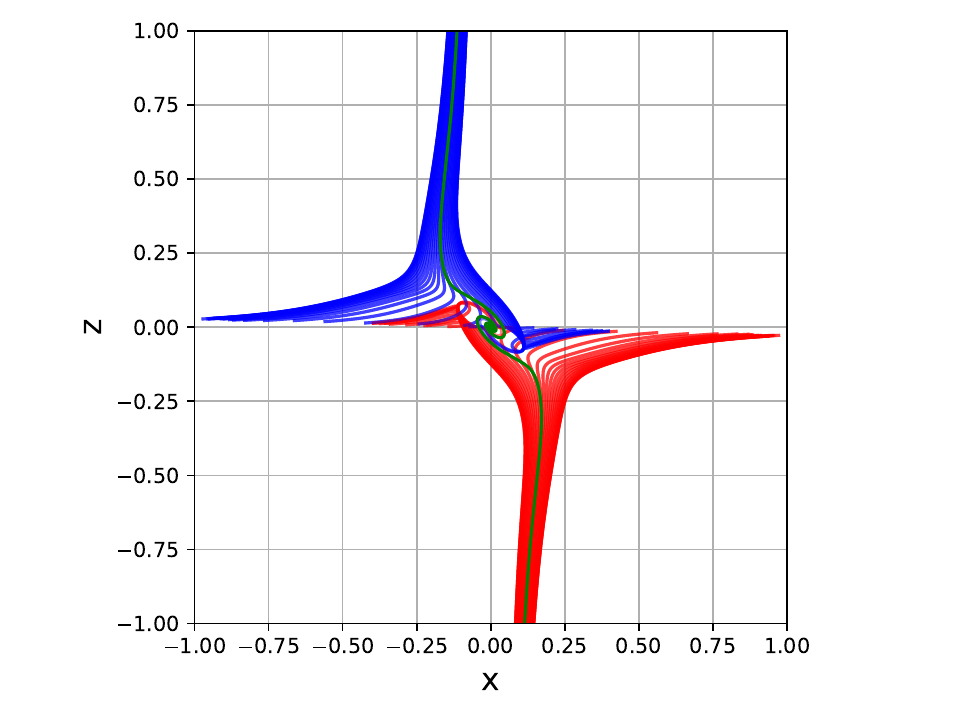}}
 \end{subfigure}
\centering
	 \begin{subfigure}[$\psi=0.1$]
 {\includegraphics[trim = 30 10 70 10, clip, width=0.32\textwidth]{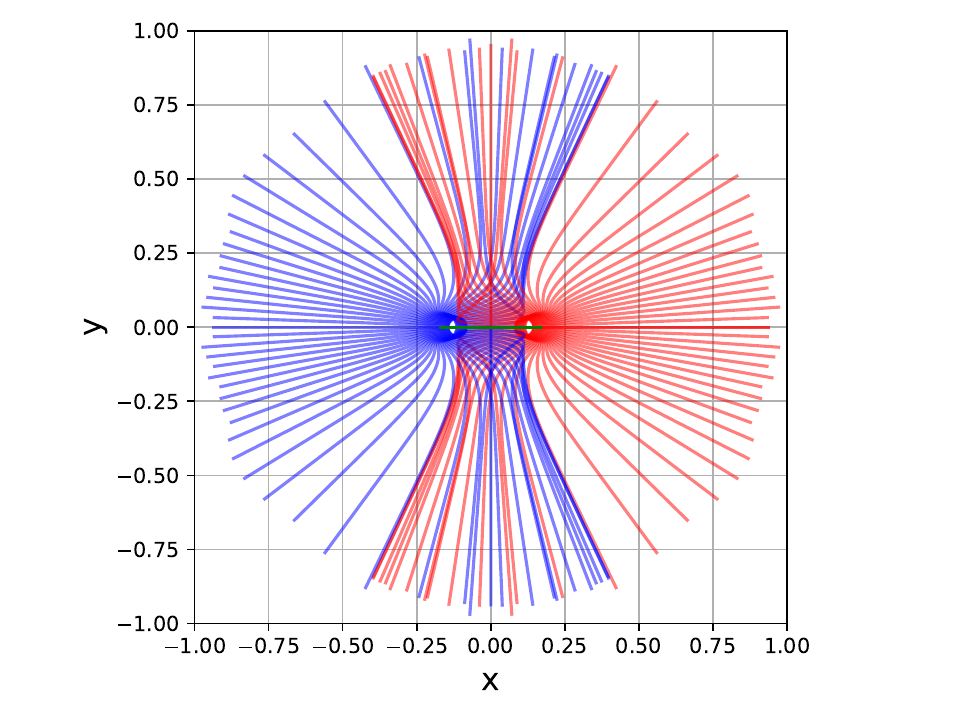}}
 \end{subfigure}
\centering
	\centering
 \begin{subfigure}[$\psi=0.1$]
 {\includegraphics[trim =  70 0 75 30, clip, width=0.32\textwidth]{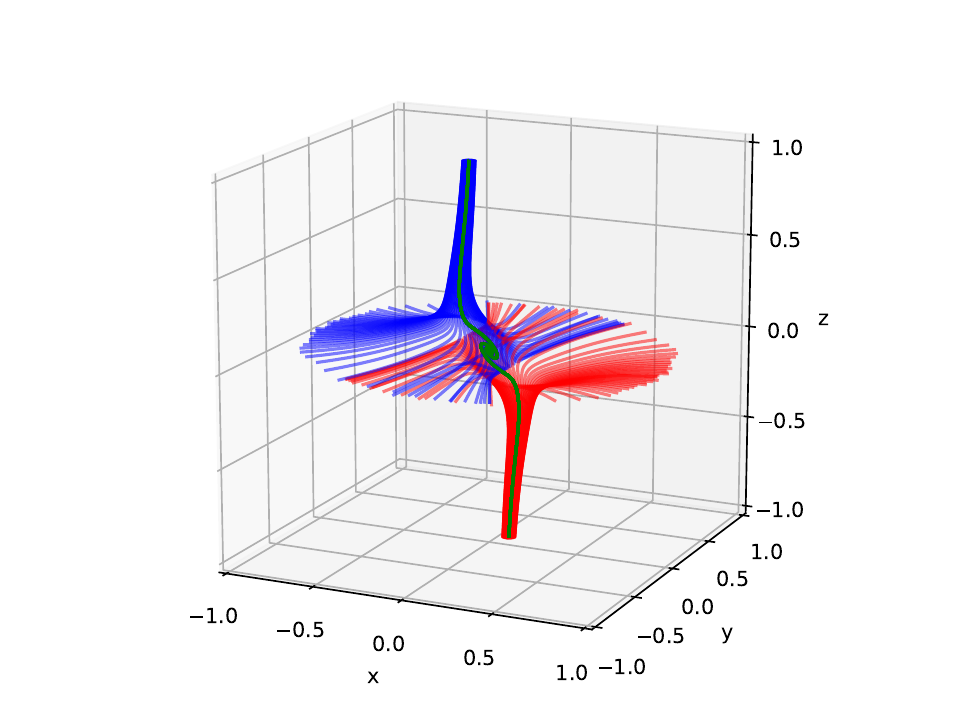}}
 \end{subfigure}
\centering
	\caption{Traced magnetic field lines for the initial condition, Equations (\ref{equation=X-point}) - (\ref{eq:potential}). The green line represents the  {\it{spine} of the null point}, while the blue and red lines represent the fan plane traced from the upper and lower boundaries, respectively. Panels (d)-(f) represent the initial condition for the baseline case, which extends \cite{Thurgood2017}. Panels (a)-(c) present the initial condition for our smallest perturbation amplitude ($\psi=0.01$) and panels (g)-(i) for the largest perturbation amplitude considered ($\psi=0.1$).}
	\label{fig:initial-field}
\end{figure*}

With regards to the the perturbation to the initial state $\mathbf{A}^\prime$, Equation (\ref{eq:potential}), the coefficients $\sigma$ and $\psi$ are constants that can be chosen to change the amplitude and spread of the initial perturbation. In this paper, we set $\sigma = 0.21$ and we vary $\psi$, which represents the initial perturbation amplitude, such that $\psi = 0.01$, $0.025$, $0.05$, and $0.1$ (weakest amplitude to strongest): 
\begin{itemize}
    \item {Figure \ref{fig:initial-field}a -   \ref{fig:initial-field}c shows that a small perturbation value (here $\psi=0.01$) which creates an initial condition with almost no bending in the spine.}
\item{The case where $\psi = 0.05$ reproduces the setup described by \citet{Thurgood2017}, as illustrated in Figure \ref{fig:initial-field}d - \ref{fig:initial-field}f, and we refer to this as our baseline simulation.}
\item{he bending of the spine due to our choice of initial condition gradually increases with increasing $\psi$, as seen in Figure \ref{fig:initial-field}g -   \ref{fig:initial-field}i, which displays the largest initial perturbation we consider, namely the simulation for $\psi = 0.1$. Here, the $\mathbf{B}^\prime$ amplitude increases significantly creating a bending of the spine and a twist around the null point.  }
\end{itemize}
We do not show figures corresponding to $\psi=0.025$ since these are very similar to our results between $\psi = 0.01$, i.e. Figure \ref{fig:initial-field}a -   \ref{fig:initial-field}c, and $\psi = 0.05$, i.e. Figure \ref{fig:initial-field}d -   \ref{fig:initial-field}f.

\subsection{Boundary conditions and domain setup}

To ensure adequate resolution in the region of primary interest (i.e. around the null), a stretched grid was employed that provided finer resolution near the null point and coarser resolution in the outer regions. The grid stretching was implemented using a hyperbolic tangent function, which smoothly transitions the grid spacing growth rate from 0 to 7\%. The grid was divided into distinct regions: a central cube with constant grid spacing and outside the cube where it gradually stretches up to the boundary.
%, using a hyperbolic tangent profile for the growth rate. 
The computational domain spans a total volume of $400^3$ grid points, with the domain extending from $-55 \leq x, y, z \leq 55$. The uniform central cube spans $-0.4 \leq x, y, z \leq 0.4$ with 100$^3$ points. Each simulation run took approximately 238,000 CPU hours for a single run. 

Neumann boundary conditions were imposed for the magnetic field and thermodynamic variables, enforcing a zero gradient at the boundaries. For the velocity field, boundary values were set to zero, i.e. creating a reflecting boundary, but crucially a sponge boundary condition was implemented in the far-field region at a radius of $r = 10$ to minimize wave reflections from the computational boundary. This condition was designed to attenuate outgoing waves by introducing a damping mechanism. The implementation follows the methodology described in \citet{bodony_2006sponge}, where a source term is incorporated into the continuity, momentum, and energy equations to remove perturbations from the initial state continuously. The modified governing equations are expressed as follows:
\begin{eqnarray}
\frac{D\rho}{D t} &=& - \rho \nabla \cdot \mathbf{v} + \xi (\rho - \overline{\rho}), \\
\rho \frac{D\mathbf{v}}{D t} &=& (\nabla \times \mathbf{B} ) \times \mathbf{B} - \nabla p + \mathbf{F}_{visc} + \rho \xi (\mathbf{v} - \overline{\mathbf{v}}) , \\
\rho\frac{De}{D t} &=& - p\nabla \cdot \mathbf{v} + \eta|\mathbf{j}|^2+  Q_{visc}+ \rho \xi (e - \overline{e}) .
\label{eq:sponge}
\end{eqnarray}
In these equations $\xi$ is the sponge coefficient that governs the strength of the damping. The sponge coefficient is defined as:
\begin{equation}
    \xi(r) = \alpha_{\text{sponge}} \left( \frac{r - r_\text{start}}{r_\text{end} - r_\text{start}} \right)^{h}, 
\end{equation}
where $\alpha_{\text{sponge}} =- 60 $ represents the damping amplitude, $r_\text{start}=10$ and $r_\text{end}=55$ mark the spatial extent of the damping region, $h = 4$ is the exponent controlling the smoothness of the damping function, and $r$ is the radial coordinate.
The sponge damping mechanism can be efficiently integrated into the solver with minimal modifications to the system of equations (\ref{eq:mass})-(\ref{eq:energy}) in discrete form by appending its contribution to the right hand side (RHS) at the conclusion of each time step. The updated variables are computed as follows:
\begin{equation}
    \rho^{n+1} = {\rm{RHS}}_{\rho}^{n}  + \Delta t \xi (\rho^{n} - \overline{\rho}),
\end{equation}
\begin{equation}
    \mathbf{v}^{n+1} = {\rm{RHS}}_{v}^{n} + \Delta t \xi (\mathbf{v}^{n} - \overline{\mathbf{v}}),
\end{equation}
\begin{equation}
    e^{n+1} = {\rm{RHS}}_{e}^{n}  + \Delta t \xi (e^{n} - \overline{e}).
\end{equation}
In this equation, the superscript $n$ denotes the current time step, and $\Delta t$ is the time step size. This approach ensures that the sponge layer effectively attenuates outgoing waves while maintaining numerical stability and accuracy.
This damping function smoothly attenuates the velocity, density and energy as they approach the domain boundary, ensuring stability in numerical simulations.
The sponge boundary condition was selected because, for hydrodynamics cases, it yields better results compared to the characteristic boundary condition. It also grants results as good as a perfect match layer, while being easier to implement \citep{zjwang_2010sponge}.

\section{Results} \label{sec:results}

\subsection{Three-dimensional oscillatory reconnection evolution} \label{sec:evolution}
Let us consider the baseline case with $\psi = 0.05$, where the perturbation amplitude corresponds to the scenario studied by \cite{Thurgood2017}. The simulation is initialized with a perturbation in the magnetic field potential, as described by Equations (\ref{eq:perturbation}) and (\ref{eq:potential}). This perturbation bends the null spine and induces a disturbance in the fan plane, as illustrated in Figures \ref{fig:initial-field}d - \ref{fig:initial-field}f. 

\begin{figure*}[p]
	\centering
    \includegraphics[width=0.97\textwidth]{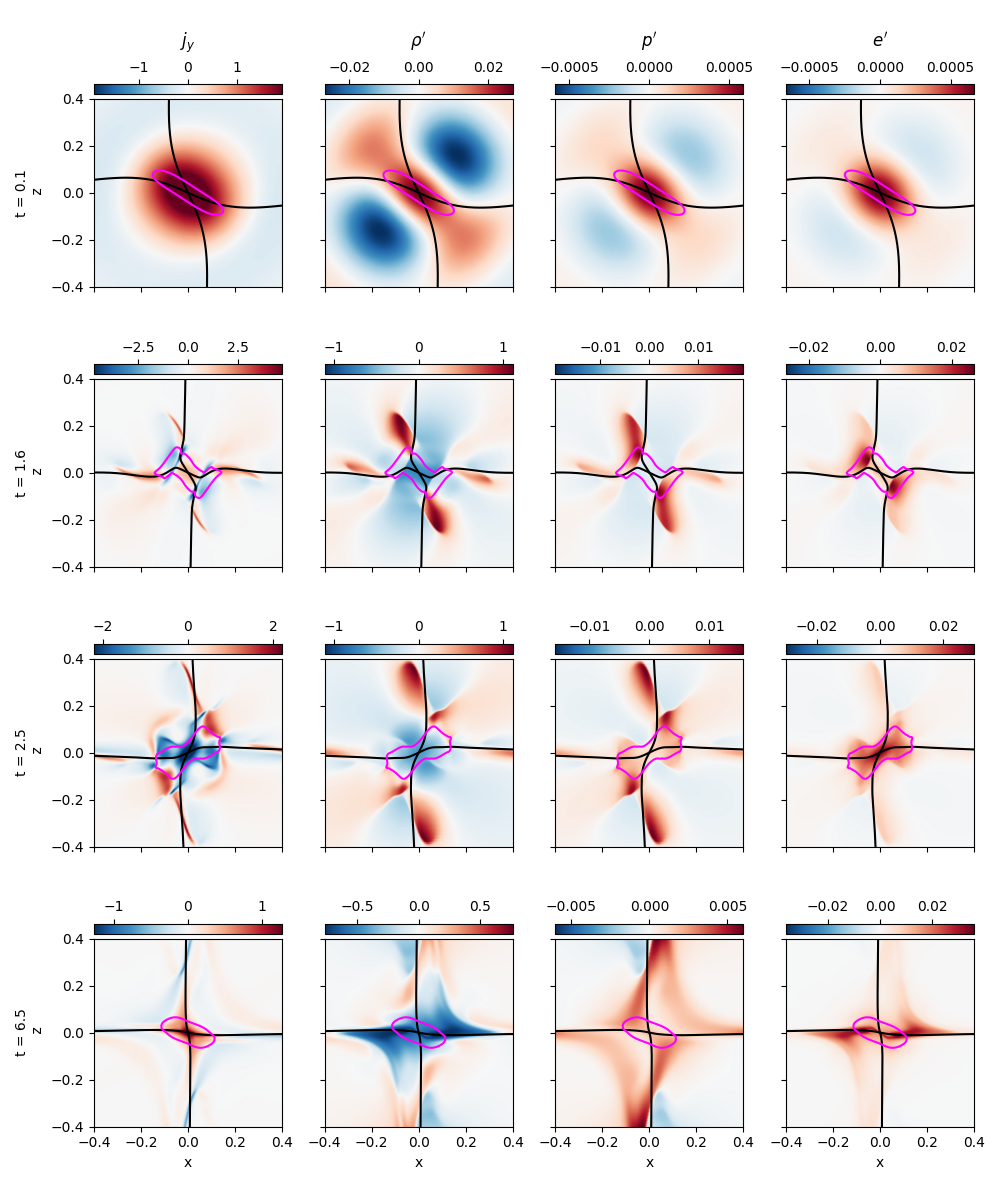}
	\caption{Contour plots in the $y=0$, $xz-$plane displaying the time evolution of the $j_y$, $\rho^\prime$, $p^\prime$, and $e^\prime$ for the baseline case, $\psi=0.05$, between $t=0.1$-6.5. The magenta lines represent the equipartition layer, and the black magnetic field lines indicate the magnetic skeleton where the $z-$axis is the spine and $x-$axis corresponds to the relevant field lines that make up the fan plane. Note that the saturation scale varies between subfigures to show the detail.}
	\label{fig:OR-evolution}
\end{figure*}

\begin{figure*}[p]
	\centering
	\includegraphics[width=0.99\textwidth]{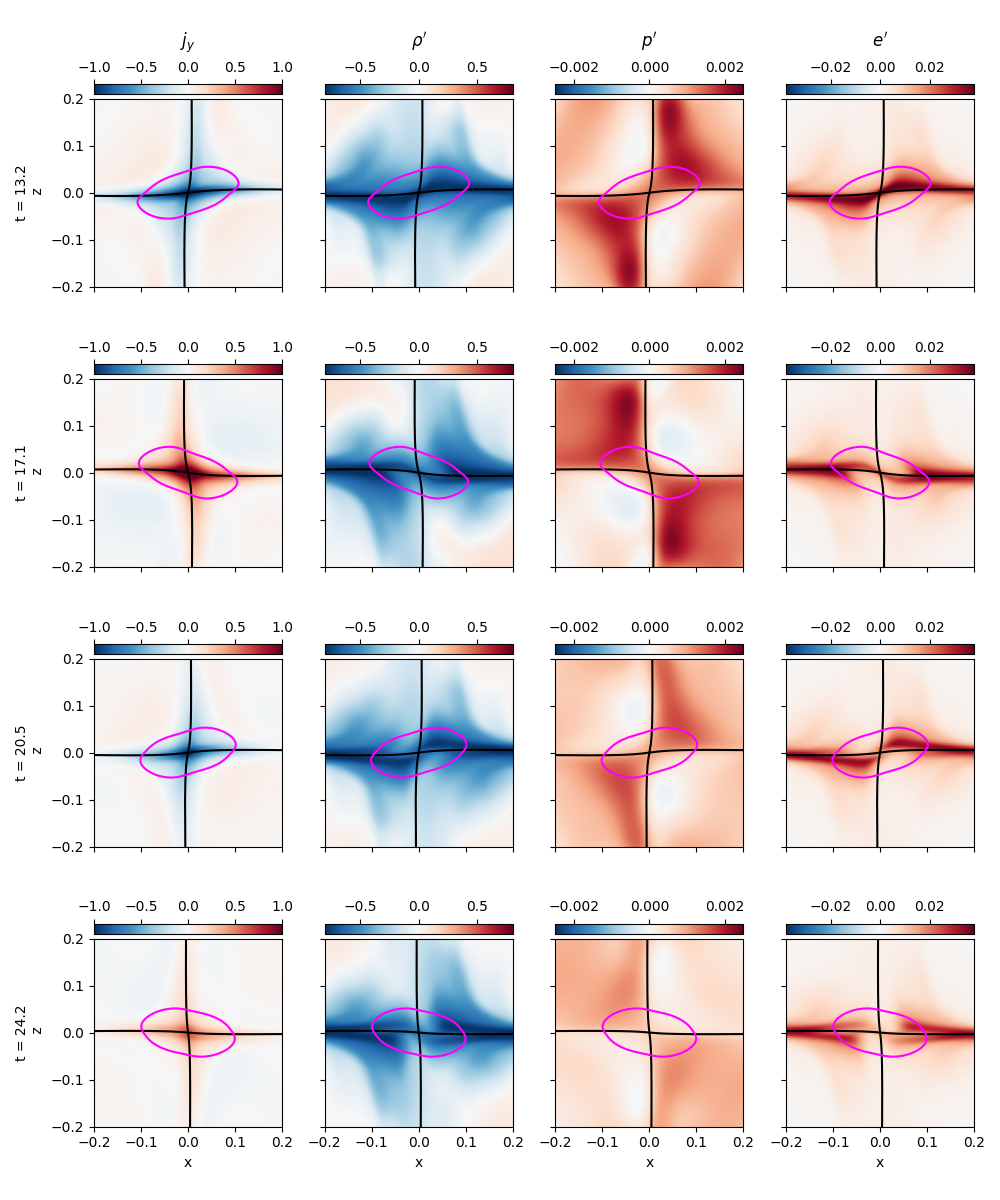}
	\caption{Contour plots in the $y=0$, $xz-$plane displaying the time evolution of the $j_y$, $\rho$, $p^\prime$, and $e^\prime$ for the baseline case, $\psi=0.05$ for later time steps. The magenta and black lines have the same meaning as in Figure \ref{fig:OR-evolution}.
    }
	\label{fig:OR-evolution-later-cycles}
\end{figure*}

\begin{figure*}[htb]
	\centering
	\includegraphics[trim = 18 230 20 0, clip,width=0.98\textwidth]{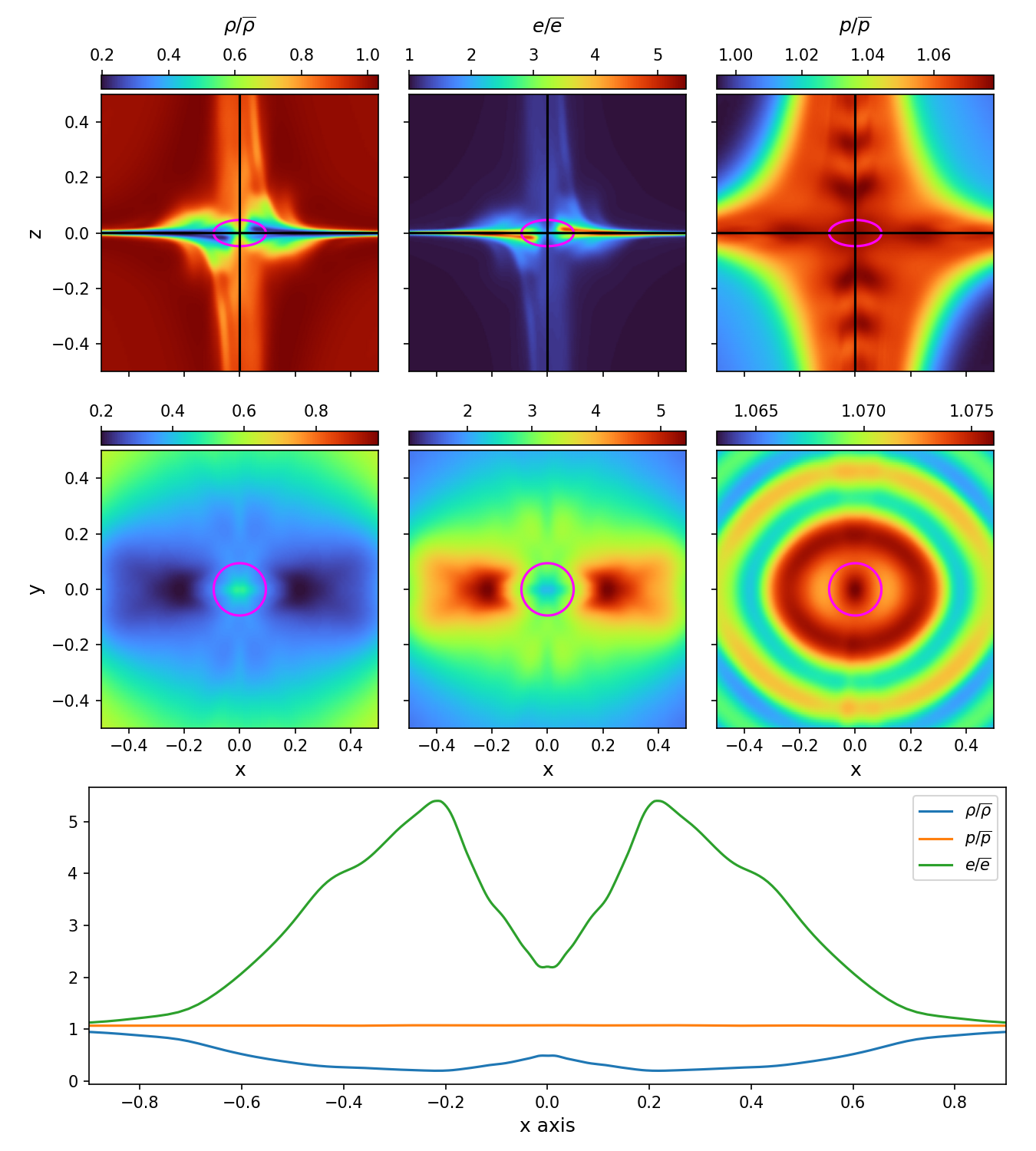}
	\caption{Contour plots or $\rho/\overline{\rho}$, $p/\overline{p}$, and $e/\overline{e}$ normalized by their initial state for the $\psi=0.05$ case at $t=60$. The magenta lines represent the equipartition layer. Top row shows the $y=0$, $xz-$plane, where the black magnetic field lines indicate the magnetic skeleton, where the $z-$axis is the spine and $x-$axis corresponds to the fan plane. Bottom row shows the $z=0$, $xy-$plane.}
	\label{fig:OR-end}
\end{figure*}

The evolution of the perturbation fields for $j_y^\prime = j_y -  \overline{j_y}$, $\rho^\prime = \rho - \overline{\rho}$, $p^\prime = p - \overline{p}$, and $e^\prime = e - \overline{e}$ at $y = 0$ is illustrated in Figure \ref{fig:OR-evolution}.  Here, the perturbation of a variable $f$ is defined as $f^\prime = f(x, y, z, t) - f(x, y, z, 0)= f - \overline{f}$, representing the deviation from the initial state. Furthermore, since our initial state is potential, $\overline{j_y}=0$ and so $j_y=j_y^\prime$. Also in the plots, the magenta line denotes the wave equipartition layer, where the sound speed $v_s$ is equal to Alfvén speed $v_a$, while the black lines represent the magnetic skeleton obtained by integrating the field lines near the null with a seed at point at $(0,0,\pm0.001)$ in order to trace the spine and $(\pm0.001,0,0)$ to trace the field lines corresponding to the fan in the $y=0$, $xz-$plane. The initial perturbation generates a localized spherical disturbance seen in $j_y$ that propagates toward the null point, as seen in Figure \ref{fig:OR-evolution} at $t = 0.1$. This spherical disturbance collapses at the null, producing outgoing perturbations along the spine and fan that are evident in $\rho^\prime$, $p^\prime$ and $e^\prime$ at $t = 1.6$ and $t = 2.5$, manifested as red blobs propagating along the spine. Here, the current sheet $j_y$ is oriented at approximately $-45^\circ$, and we refer to this as the current sheet being in \lq{orientation one}\rq{}.

Additionally, the reorientation of the magnetic field lines between $t = 1.6$, $t = 2.5$, and $t = 6.5$ characterizes the first cycle of the oscillatory reconnection mechanism. At $t = 2.5$,  the current sheet $j_y$ is oriented at approximately $+45^\circ$, and we refer to this as the current sheet being in \lq{orientation two}\rq{}. Then at $t = 6.5$,  the current sheet $j_y$ has returned back to an orientation of approximately $-45^\circ$, i.e. the current sheet has return back to \lq{orientation one}\rq{}. The perturbation in internal energy reveals that the initial pulse heats the region surrounding the null point at $t = 0.1$. However, at $t = 1.6$ and $t = 2.5$, further heating occurs due to reconnection jets. By $t = 6.5$, the energy perturbation rises, and the increase in internal energy remains localized around the fan plane, as heat conduction is not considered in this simulation. (See \cite{Karampelas2022a} for a consideration of heat conduction around a 2D null.)

Finally, propagating perturbations in $j_y$ along the spine are observed at $t = 2.5$ and $t = 6.5$, which appear to be synchronized with the oscillatory reconnection cycle. These features highlight the dynamic interplay between magnetic reconnection and wave propagation near the null point.

Figure \ref{fig:OR-evolution-later-cycles} presents several later time steps illustrating the decay of the oscillatory reconnection cycles. At $t = 13.2$, {the current sheet $j_y$ is at
\lq{orientation two}\rq{}.} There is a significant drop in density near the null point, along with heating observed in $e^\prime$ along the current sheet, reconnection jets, and the fan plane. Additionally, the pressure perturbation increases and is considerably higher along the reconnection jets.

At $t = 17.1$, we have reached a later cycle in the oscillatory reconnection phenomenon and the current sheet is now back in \lq{orientation one}\rq{}. Similar to the situation at $t = 13.2$, there is again a significant drop in density near the null point and fan plane, along with heating ($e^\prime$) in the same region. The increase in pressure perturbation is also reoriented to $-45^\circ$ (`orientation one').

\begin{figure*}[t]
	\centering
	\includegraphics[width=0.99\textwidth]{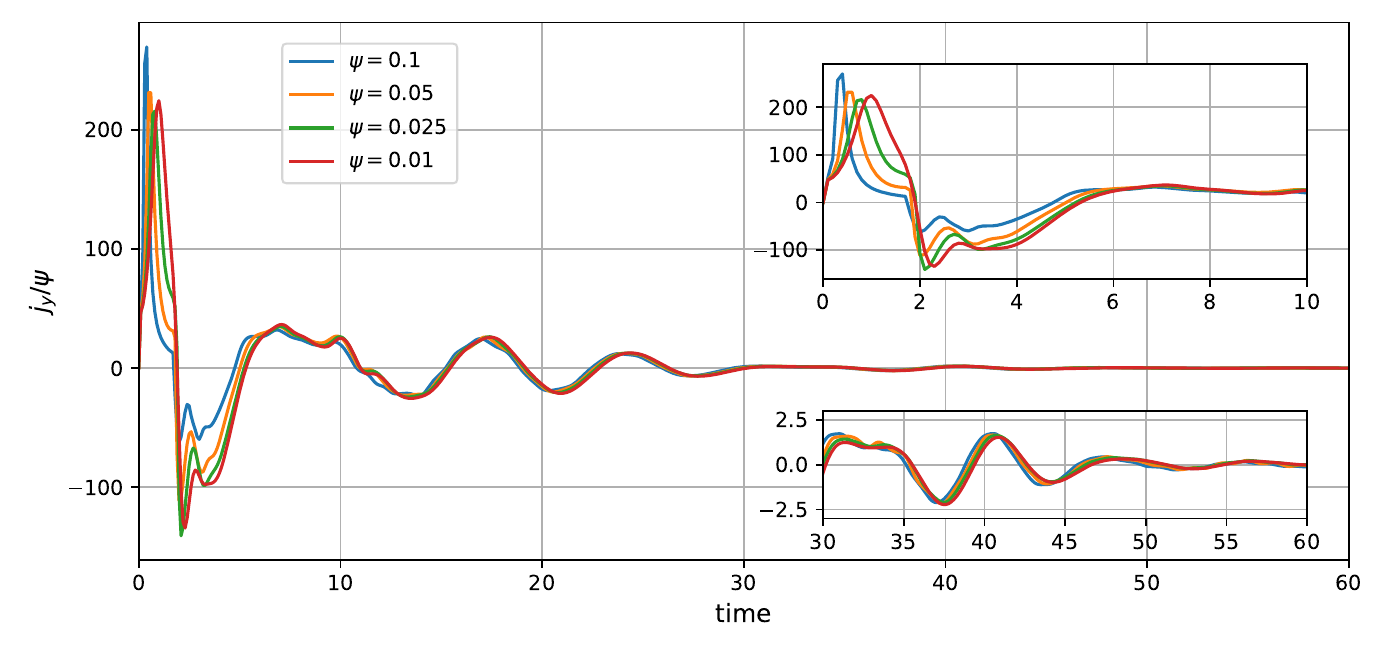}
%	\caption{The $y$ component of current density measured at the null point, normalized by the perturbation amplitude, $j_y(0,0,0,t)/\psi$, for each simulation.}
\caption{The evolution of $j_y(0,0,0,t)$, measured at the null point and normalized by the initial perturbation amplitude, $\psi$,  for $\psi = 0.01$, $0.025$, $0.05$ and $0.1$. Insets show same evolution but over subsets of time (to show further detail).}
        \label{fig:jy-oscillation}
\end{figure*}

The time points $t = 20.5$ \lq{orientation two}\rq{} and $t = 24.2$ \lq{orientation one}\rq{} follow the same patterns observed at $t = 13.2$ and $t = 17.1$, respectively, demonstrating a reorientation of the current sheet due to oscillatory reconnection. Although the patterns at $t = 20.5$ and $t = 24.2$ are similar to those at $t = 13.2$ and $t = 17.1$, it is evident that the amplitude of $p^\prime$ is decaying, as is $j_y$. This will be further explored in $\S\ref{sec:jy-at-null}$  and Figure \ref{fig:jy-oscillation}.

\begin{figure*}[t]
	\centering
	\includegraphics[trim = 0 20 0 20, clip, width=0.99\textwidth]{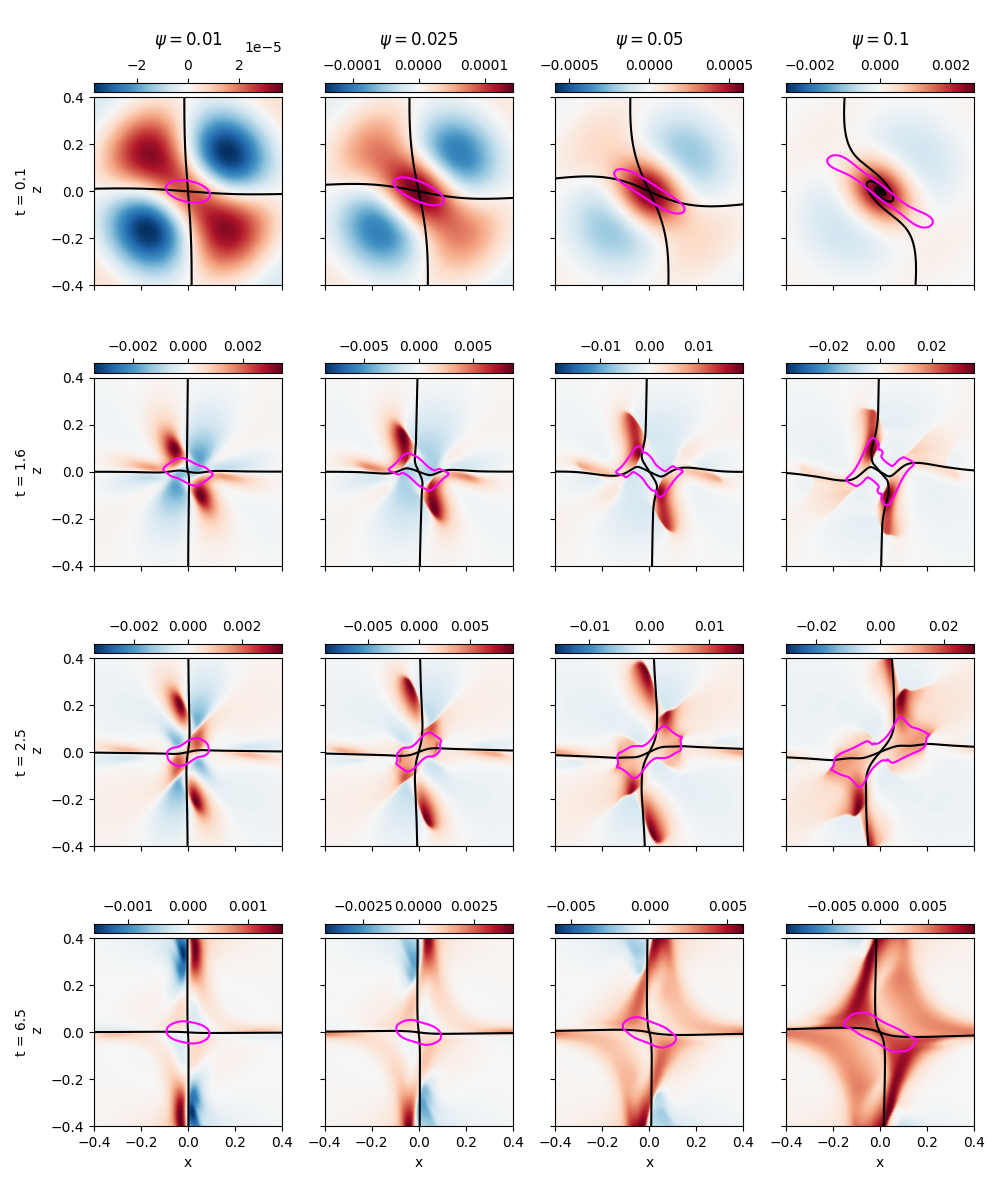}
	\caption{Contour plots in the $y=0$, $xz-$plane displaying the time evolution of the $p^\prime = p - \overline{p}$ contours for simulations $\psi=0.01$, $\psi=0.025$, $\psi=0.05$, and $\psi=0.1$ (corresponding to the first, second, third and fourth columns, respectively). Note that the third column of Figure \ref{fig:OR-psi} is identical to that of the third column of Figure \ref{fig:OR-evolution}, but is replicated here to ease comparison. The magenta lines represent the equipartition layer, and the black magnetic field lines indicate the magnetic skeleton, where the $z-$axis is the spine and $x-$axis corresponds to the fan plane. Note that the saturation scale varies between subfigures to show the detail.}
	\label{fig:OR-psi}
\end{figure*}

\subsubsection{End of simulation} \label{sec:end}

Figure \ref{fig:OR-end} displays $\rho/ \overline{\rho}$, $p / \overline{p}$ and $e / \overline{e}$ for the case of $\psi=0.05$ at $t=60$. Here, there is no visible bending in the spine at the end of the simulation, indicating that the system has relaxed to a state close to its original equilibrium. The density plots reveal a significant decrease in density along the fan plane (at $t=60$), with values ranging from three to five times smaller than the initial state. Similarly, internal energy increases three to five times in the fan plane. The heat remains trapped in the fan plane due to our model's lack of heat conduction. The pressure shows an approximately 6\% increase from the initial base state near the null, where $p/\overline{p}=1.06$. Additionally, small pressure oscillations can be observed propagating along both the spine and the fan plane. Furthermore, $j_y(0,0,0,t=60)=0$ and so, since there is no further oscillatory reconnection cycles,  we consider $t=60$ as the end of our simulation.

\subsection{Influence of  initial perturbation amplitude, $\psi$} \label{sec:evolution-drivers}

Figure \ref{fig:OR-psi} presents a comparative analysis of the evolution of the pressure perturbation, $p^\prime$, during the first reconnection cycle, i.e. the time taken to go from an \lq{orientation one}\rq{} current sheet, through an \lq{orientation two}\rq{} current sheet, and then revert back to an \lq{orientation one}\rq{} current sheet.

%first period, across all simulations. 

At $t = 0.1$, the initial spherical perturbation propagates toward the null point. The simulation with $\psi=0.01$ shows the least deformation of the spine and exhibits the weakest amplitude of perturbation compared to the other cases. Despite significant variations in the initial spine configurations among simulations, the spatial distribution of $p^\prime$ remains similar, with differences only in amplitude.

By $t = 1.6$, the first reconnection event occurs in all simulations, accompanied by the formation of a high-pressure region within the reconnection jets. The morphology of the equipartition layer evolves from a nearly elliptical shape for $\psi=0.01$ to a more distorted configuration for $\psi=0.1$. The geometry of the equipartition layer varies considerably, while the overall behavior of the system remains consistent across all four cases, but presenting different orders of magnitude across the pressure levels.

At $t = 2.5$, magnetic field line reorientation and jet formation are observed universally. Wave propagation along the spine becomes evident, with the $\psi=0.1$ case showing an earlier departure of perturbations compared to simulations with lower values of $\psi$.

By $t = 6.5$, the pressure distributions evolve to similar patterns across all cases. At this stage, the equipartition layer adopts a more elliptical shape across all simulations after the first cycle of oscillatory reconnection. Note there is a small difference in the evolution of pressure between the four amplitude cases, due to stronger initial perturbations increasing magnetic tension, slightly accelerating the initial reconnection cycle relative to the lower amplitude cases.

\subsubsection{Current density evolution at the null point}\label{sec:jy-at-null}

%Figure \ref{fig:jy-oscillation} illustrates the signature of oscillatory reconnection, as measured in the current density component $ j_y $ at the null point.

The evolution of the current density at the null point is a key signature of oscillatory reconnection.  In our simulations, the position of the null point was identified and tracked over time using the null point identification algorithm described in \cite{haynes_trilinear_2007}. It was found that the null point remains stationary and is consistently located at $x = y = z = 0$ throughout all simulations. This location of the null point holds true for all perturbation amplitudes, $\psi$, due to the symmetric nature of the applied perturbation. Thus, we can measure the evolution of the current density at the null point via $j_y(0,0,0,t)$ and this can be seen in Figure \ref{fig:jy-oscillation}.

Figure \ref{fig:jy-oscillation} reports on the signature of oscillatory reconnection in a three-dimensional (3D) simulation extended up to 60 Alfvén time scales ($t=60$ in our non-dimensional variables)
%, namely $j_y(0,0,0,t)$ 
for initial perturbation amplitudes $\psi = 0.01$, $0.025$, $0.05$, and $0.1$). This is presented as $j_y / \psi$ in order to best compare the four cases.

\begin{figure}[b]
	\centering
\includegraphics[trim = 10 0 0 0, clip, width=0.98\columnwidth]{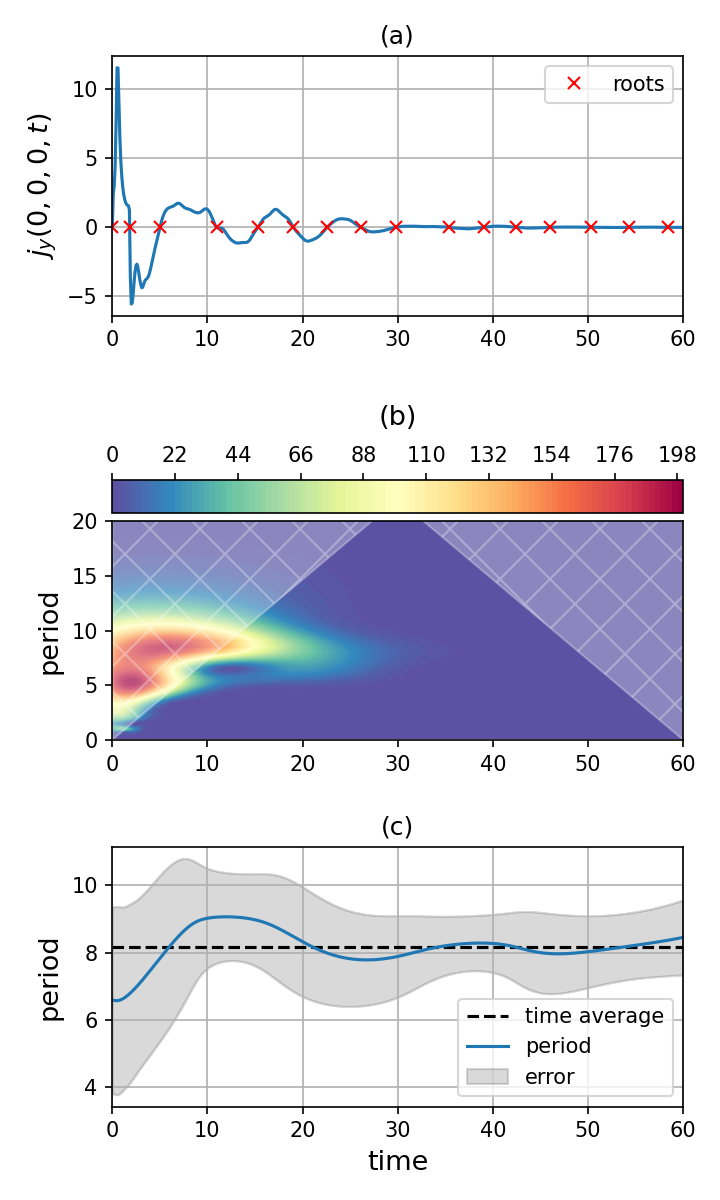}
	\caption{Period extraction for the baseline case. (a) Current density at the null point, $j_y(0,0,0,t)$. (b) $j_y(0,0,0,t)$ wavelet power spectrum, where the white-hatched region represents the cone of influence. (c) Dominant period obtained from the wavelet power map at each time step by applying a Gaussian fitting, where the blue line shows the Gaussian mean and the shaded area the estimated error using the Gaussian standard deviation. The dashed line show the average period obtained from $t=$0 to 60. }
	\label{fig:jz-wavelet}
\end{figure}

This damped oscillatory behavior when normalized by the perturbation amplitude is invariant, as demonstrated in \citet{Karampelas2022b} and \citet{schiavo2024energymap}. Furthermore, it appears independent of the initial perturbation, as discussed in \cite{Karampelas2022b}.
In the 3D simulations presented here, the $j_y(0,0,0,t)/\psi$ oscillates around a zero value, in contrast to the 2D case which stabilizes to a non-zero-value \citep{McLaughlin2009}.
The initial peak in $j_y(0,0,0,t)$ occurs at $t=0.4$, 0.5, 0.8 and 1 for $\psi=0.1$, 0.05, 0.025 and 0.01, respectively. The strongest perturbation initiates the reconnection cycle earlier than the cases with smaller $\psi$. As discussed in Section 3.2, stronger perturbations increase the magnetic tension, which also raises the Alfvén speed near the null point. This causes the perturbation to reach the null earlier, resulting in different times for the first peak in $j_y(0,0,0,t)$. The first peak is larger than the subsequent ones because it is dominated by the influence of the spherical perturbation (as shown in Figure \ref{fig:OR-evolution} at $t=0.1$), which collapses at the null, creating an overshoot in the current density. This effect was also reported in \cite{McLaughlin2009}.

%%%%%%%%%%%%%%%%%%%%%%%%%%%%%%%%%%%%%%%%%%%%%%

%\subsubsection{Characterization of the oscillation period} \label{sec:period}

\subsubsection{The periodicity of three-dimensional oscillatory reconnection} \label{sec:period}

To characterize the oscillation period derived from $j_y(0,0,0,t)$ a continuous wavelet analysis is performed using a Morlet wavelet with a  central frequency of 6.
Figure \ref{fig:jz-wavelet}a displays the $j_y(0,0,0,t)$ signal and its roots.
Figure \ref{fig:jz-wavelet}b presents the wavelet power diagram.
Figure \ref{fig:jz-wavelet}c shows the instantaneous periods resulting from the wavelet analysis.

%$\omega_0 = 6$.

%Figure \ref{fig:jz-wavelet}a displays the $j_y(0,0,0,t)$ signal and its roots. Figure \ref{fig:jz-wavelet}b presents the wavelet power diagram, where red indicates high power and blue corresponds to low power. A white-hatched region represents the Cone-of-Influence. Figure \ref{fig:jz-wavelet}c shows the dominant period obtained by fitting a Gaussian function to the dominant period at wavelet power map each time step. The instantaneous period is given by the Gaussian mean, show in blue, and and gray areas shows the estimated error using the gaussian standard deviation. 

The extracted periods for initial perturbation amplitudes $\psi = 0.01$, $0.025$, $0.05$, and $0.1$ are summarized in Table \ref{tab-period-wavelet} for multiple time series intervals: $t = 0-60$, $t = 5-60$ and $t = 20-60$, in order to analyze the influence of the initial transient on the period. For the entire time series ($t = 0-60$), the average period ranges from 8.1 to 8.3, with an estimated error between 1.3 and 1.4. When analyzing the time series for $t = 20-60$, the period becomes more consistent across the four simulations, stabilizing around 8.1, and the uncertainty decreases as the initial transient is excluded from the analysis.

%\begin{table}[htb]\centering
\begin{table}[b]\centering
\begin{tabular}{lcccc}\hline 
\textbf{Time}& \multicolumn{4}{c}{\textbf{Average period}} \\ 
      &$\psi$= 0.01   & $\psi$= 0.025   & $\psi$= 0.05   & $\psi$= 0.1  \\\hline \hline 
0-60  & 8.2 $\pm$ 1.3 & 8.3 $\pm$ 1.3 & 8.2 $\pm$ 1.4 & 8.1 $\pm$ 1.4      \\
5-60  & 8.3 $\pm$ 1.2 & 8.1 $\pm$ 1.4 & 8.3 $\pm$ 1.3 & 8.2 $\pm$ 1.2      \\
20-60 & 8.2 $\pm$ 1.1 & 8.1 $\pm$ 1.1 & 8.1 $\pm$ 1.1 & 8.0 $\pm$ 1.1     \\ \hline 
\end{tabular}
	\caption{Average period of the $j_y(0,0,0,t)$ oscillation  extracted from wavelet analysis, for initial perturbation amplitudes $\psi = 0.01$, $0.025$, $0.05$ and $0.1$.}
	\label{tab-period-wavelet}
\end{table}

%%%%%%%%%%%%%%%%%%%%%%%%%%%%%%%%%%%%%%%%%%%%%%%%%%%%%%%%%%%%%%%%%%%%

%\begin{figure}[htb]
\begin{figure}[t]
	\centering
	\includegraphics[trim = 15 0 0 0, width=0.99\columnwidth]{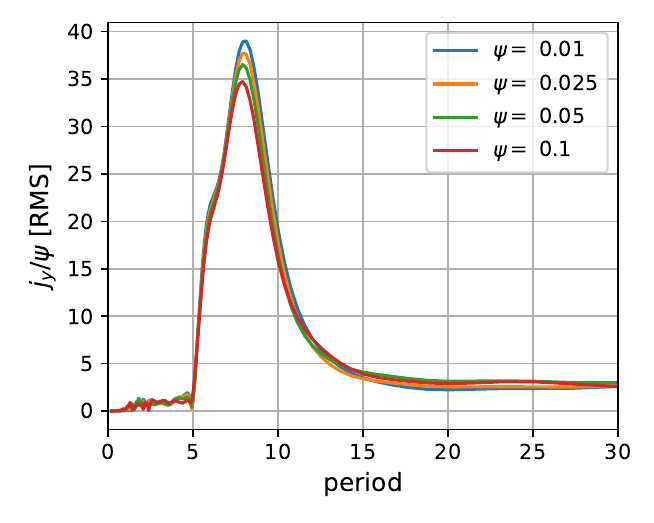}
	\caption{Fourier spectra for $j_y(0,0,0,t)/\psi$, for initial perturbation amplitudes $\psi = 0.01$, $0.025$, $0.05$ and $0.1$.}
	\label{fig:jy-fft}
\end{figure}

\begin{figure*}[t]
	\centering
	\includegraphics[ width=0.99\textwidth]{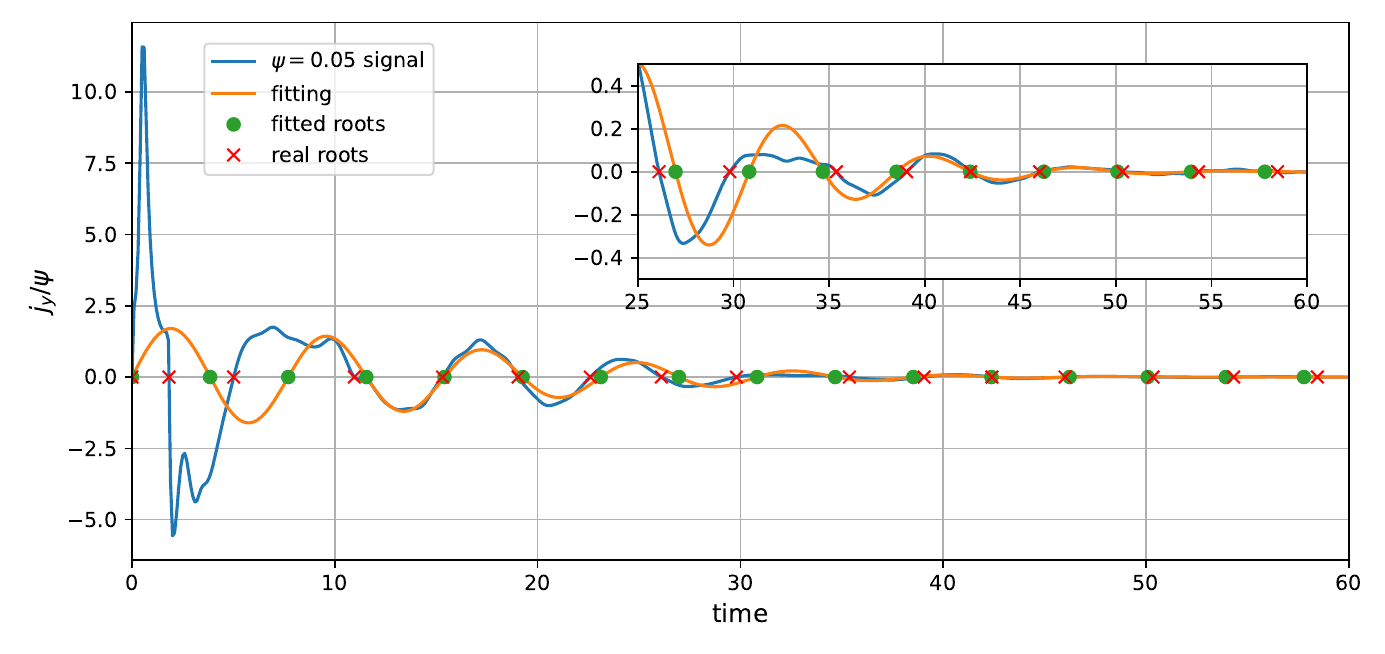}
	\caption{Current density $j_y(0,0,0,t)$ measured at the null point compared with the best-fit functions $g_2$ and $f_1$ from Equations (\ref{eq:f_1}) and (\ref{eq:g2}) respectively.}
    \label{fig:jy-fitting}
\end{figure*}

Figure \ref{fig:jy-fft} displays the Fourier spectra of $j_y/\psi$ for each simulation case. The spectra is invariant, with minor differences in amplitude and a consistent dominant frequency. A 
fitting of a Gaussian function was applied to the dominant period, and the dominant periods are presented in Table \ref{tab-period-FFT}. The extracted period from the Fourier analysis agrees with the wavelet analysis; however, the Fourier spectra results have a larger estimated error.

%\begin{table}[htb]\centering
\begin{table}[htb]\centering
\begin{tabular}{lcccc}\hline 
\textbf{Time} & \multicolumn{4}{c}{\textbf{Average period}} \\ 
 &$\psi$= 0.01   & $\psi$= 0.025   & $\psi$= 0.05   & $\psi$= 0.1  \\\hline \hline 
0-60  & 8.1 $\pm$ 1.7 & 8.0 $\pm$ 1.6 & 8.0 $\pm$ 1.6 & 8.0 $\pm$ 1.6      \\  \hline 
\end{tabular}
%	\caption{Average $j_y(0,0,0,t)$ oscillation period extracted from Fourier spectra. }
	\caption{Average period of the $j_y(0,0,0,t)$ oscillation extracted from Fourier spectra, for initial perturbation amplitudes $\psi = 0.01$, $0.025$, $0.05$ and $0.1$.}
   	\label{tab-period-FFT}
\end{table}

\subsection{Analytical form of $j_y(0,0,0,t)$}\label{sec:jy-modeling}

$\S\ref{sec:jy-at-null}$ reported that $j_y(0,0,0,t)$ is invariant when normalized by $\psi$. %Let us now 
In this section we model $j_y(0,0,0,t)$ as the product of two functions: an oscillatory function $f(t)$ and a decaying envelope $g(t)$, such that:
\begin{equation}
j_y(0,0,0,t) = g(t)f(t) .
\end{equation}
Two candidate functions are considered to approximate the oscillatory component:
\begin{eqnarray}
f_1(t) &=& \sin\left[ \frac{2\pi}{\Lambda} (t-\phi)\right] , \label{eq:f_1}\\
f_2(t) &=& J_n(   \Omega t) ,
\end{eqnarray}
where $ \Lambda $ represents the oscillation period, $\phi$ is the phase shift, $J_n$ is the Bessel function of the first kind of order $n$, and $\Omega$ is a constant that scales time in the Bessel argument. $f_1$ was chosen to capture the constant time averaged periodicity as seen in Figure \ref{fig:jz-wavelet}c. However, the roots of $j_y(0,0,0,t)$ are not strictly periodic at early times, due to the initial transient as seen in Figure \ref{fig:jz-wavelet}a. 
Therefore, $f_2$ was considered where Bessel functions are particularly suitable choice, as they exhibit oscillatory behavior but their roots are not uniformly spaced, except asymptotically for large $t$.

The decaying envelope $g(t)$ was approximated using three different profiles:
\begin{eqnarray}
g_1(t) &=& a \exp\left(-\frac{t}{\tau_e}\right), \label{eq:g1}\\
g_2(t) &=& a \exp\left(-\frac{t^2}{2 \tau_g^2}\right), \label{eq:g2}\\
g_3(t) &=& 
\begin{cases}
\displaystyle{a \exp\left(-\frac{t^2}{2 \tau_g^2}\right)} , &  t < t_s \\
\displaystyle{b \exp\left(-\frac{t - t_s}{\tau_e}\right)} , &  t > t_s  
\end{cases} \label{eq:g3}
\end{eqnarray}
where $g_1$ represents an exponential decay, $g_2$ signifies a Gaussian decay, and $g_3$ refers to a generalized damping profile (GDP), with $\tau_e$  the exponential time scale,  $\tau_g$ the Gaussian time scale, $t = t_s$ the transition time from a Gaussian to an exponential decay for GDP, and $a$ and $b$ are constant amplitudes. For $g_3$, the continuity of both the function and its first derivative at $t = t_s$ is ensured by specifying coefficients
\begin{equation}
a = b \exp\left(\frac{t_s^2}{2 \tau_g^2}\right) \quad \rm{and} \quad \tau_g = \sqrt{t_s \tau_e} . 
\end{equation}
Equation (\ref{eq:g3}) has been used previously to model damping profiles in coronal loop oscillations \citep{pascoe_damping_2013,Nakariakov2021}.

%%%%%%%%%%%%%%%%%%%%%%%%%%%%%%%%%%%%%%%%%%%%%%%%%%

\begin{table}[b]
\centering
\begin{tabular}{lcc} \hline \hline
$f(t)$ & $R^2$ & RMS error    \\ \hline
$\displaystyle{\sin\left[ \frac{2\pi}{\Lambda} (t-\phi)\right]} $  & 0.9972   & 0.9766 \\
$J_0(\Omega t)$   & 0.9855   & 2.2134 \\
$J_1( \Omega t)$   & 0.9949   & 1.3138 \\
$J_2( \Omega t)$   & 0.9884   & 1.9754 \\ \hline
\end{tabular}
	\caption{Fitting results for $f(t)$, showing the coefficient of determination $R^2$ and the RMS error for the root positions of $j_y(0,0,0,t)$.}
	\label{tab-root-fit}
\end{table}

\subsubsection{Fitting the oscillatory component $f(t)$}\label{sec:3.3.1}

We fit the function $f(t)$ to minimize the root-mean-square (RMS) error in the root positions of $j_y(0,0,0,t)$ for the baseline case ($\psi=0.05$). This optimization allowed us to identify the parameters that best align the roots of $f(t)$ with those of $j_y(0,0,0,t)$. We assessed the RMS error and the coefficient of determination $R^2$ for each fit, and the results are summarized in Table~\ref{tab-root-fit}.

We find that $f_1(t)$ produces the lowest RMS error and the highest $R^2$ value, indicating that it provided the best fit. The Bessel function of order zero, $J_0$, also performed well, exhibiting a small RMS error and a high $R^2$. The optimal parameters obtained were $\Lambda = 7.7$ (the period), $\phi = 0$ for  the sine function $f_1(t)$, and $\Omega = 0.8$ for $J_1$. Note that the wavelet analysis gives a period of $8.2\pm1.4$ and the fit gives $\Lambda=7.7$. This difference is within the Gaussian standard deviation as shown in Figure \ref{fig:jz-wavelet}. It is worth noting that the oscillation period derived from this fitting is slightly shorter than that obtained from wavelet and Fourier analyses.

%%%%%%%%%%%%%%%%%%%%%%%%%%%%%%%%%%%%%%%%%%%%%

\subsubsection{Fitting the combined function $g(t)f(t)$}\label{sec:3.3.2}

After finding the best-fit oscillatory function $f(t)$, the combination $g(t)f(t)$ was optimized to minimize the RMS error in fitting $j_y(0,0,0,t)$ over the time interval $7.5 < t < 60$. This range was selected to exclude the initial transient phase, which could distort the fit. The coefficients for $f(t)$ were fixed to the values obtained in $\S\ref{sec:3.3.1}$ and  Table~\ref{tab-root-fit}.

The results of all six combined functions can be found in Table \ref{tab-fit-envelope}. The combination that provided the best fit was a sine function, Equation (\ref{eq:f_1}), paired with a Gaussian envelope, Equation (\ref{eq:g2}). Although combining the sine function %(Equation \ref{eq:f_1}) 
with the Gaussian derivative profile, Equation (\ref{eq:g3}), produced a comparable RMS error, it introduced extra complexity without offering significant improvement. Additionally, the Bessel function $J_0$ with a Gaussian envelope yielded a reasonable fit but resulted in a higher RMS error.

Figure \ref{fig:jy-fitting} compares $j_y(0,0,0,t)$ with the best-fit combination of $g_2(t)f_1(t)$. The fitted function closely matches the oscillatory reconnection signal, particularly in the alignment of the roots. This analysis demonstrates that the damping profile of oscillatory reconnection is best described by a Gaussian decay. The GDP, though widely used in solar physics, does not offer significant advantages for this system.
In the baseline case, the optimal values for the decaying envelope are $a=1.71$ and $\tau_g=16.03$, resulting in the following expression for the $j_y(0,0,0,t)$:
\begin{equation}
j_y(0,0,0,t) = 1.71\exp\left(-\frac{t^2}{513.92}\right) \sin(7.7 \ t) .
\end{equation}

\begin{table}[htb]
\centering
\begin{tabular}{lllll}  \hline \hline
   & \multicolumn{2}{c}{$f_1(t)$} & \multicolumn{2}{c}{$f_2(t)$} \\ 
   & $R^2$ & RMS error & $R^2$ & RMS error        \\ \hline
$g_1(t)$ & 0.8662 & 0.1701 & 0.8017  &  0.2181 \\
$g_2(t)$ & 0.8976 & 0.1488 & 0.825  &  0.204 \\
$g_3(t)$ & 0.8976 & 0.1488 & 0.8017  &  0.2181 \\ \hline
\end{tabular}
	\caption{Fitting results for $g(t)f(t)$, showing the coefficient of determination $R^2$ and the RMS error for the root positions of $j_y$. }
	\label{tab-fit-envelope}
\end{table}

%%%%%%%%%%%%%%%%%%%%%%%%%%%%%%%%%%%%%%

\subsection{Net circulation $\Gamma$}  \label{sec:circulation}

\begin{figure*}[t]
\begin{center}
 \begin{subfigure}
 {\includegraphics[trim = 10 0 0 0, clip, width=0.47\textwidth]{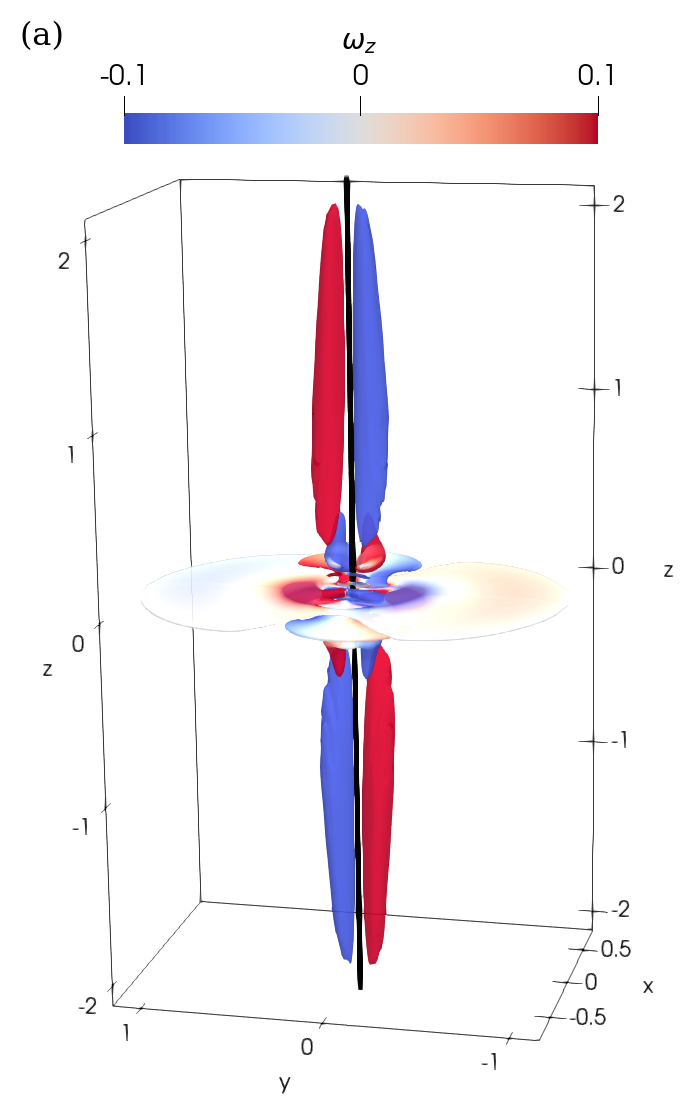}}
 \end{subfigure}
 \begin{subfigure}
{\includegraphics[trim=10 0 12 0, clip, width=0.48\textwidth]{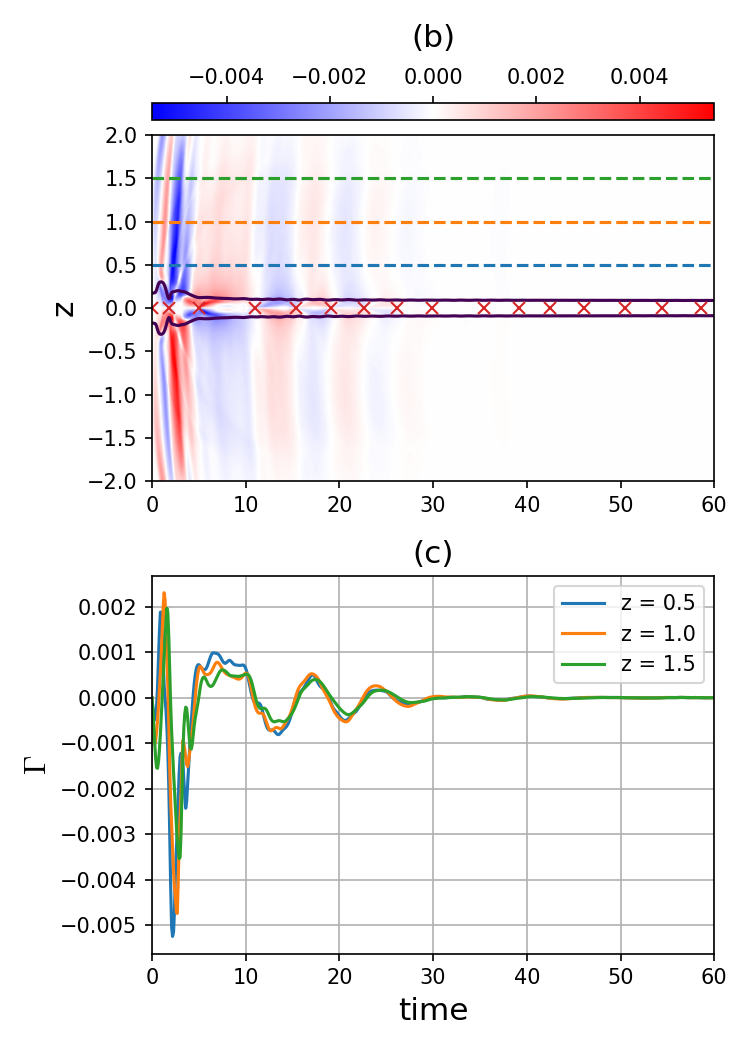}}
 \end{subfigure}
\caption{(a) Isosurfaces of Q criterion showing vortex rolls colored by $\omega_z$, where the black line represents the spine. (b)  shows a time-distance diagram of the circulation, $\Gamma(z,t)$, integrated over the $xy$-plane at every position along the $z$ axis for the $\psi = 0.05$ case. Red crosses indicate the roots of $j_y(0,0,0,t)$ and the black line represents the averaged equipartition layer.  Dashed lines correspond to $z=0.5$, $1$ and $1.5$.  (c) plots the $\Gamma (z,t)$ at three different heights, $z=0.5$, $1$ and $1.5$, taken from the dashed lines in (b).}
\label{fig:circulation-diagram}
\end{center}
\end{figure*}

To investigate the potential existence of azimuthal motions associated with torsional waves around the spine, the circulation, $\Gamma(z,t)$, was computed as a measure of net torsional motion at multiple heights. The circulation is defined as the line integral of the velocity vector field around a closed curve:
\begin{equation}
\Gamma = \oint_{C} \mathbf{v} \cdot d\mathbf{l} =
\iint_S \nabla \times \mathbf{v} \cdot d\mathbf{S} = 
\iint_S \pmb{\omega} \cdot d\mathbf{S}
\end{equation}
where $C$ is a closed curve, $S$ is the surface bounded by $C$ and $\pmb{\omega} = \nabla \times \mathbf{v}$ is the vorticity vector. 
% Figures \ref{fig:circulation-diagram}b and \ref{fig:circulation-diagram}c shows the circulation integrated over the $xy-$planes for the baseline case at different time steps. 
The integration surface, $S$, is chosen to span the region $0 <  y < 1$ and $-1 <  x < 1$, which is sufficiently large to capture vorticity oscillations in the vicinity of the spine in one side of the symmetry plane, $y=0$.
We find that $\Gamma$ is anti-symmetric about the $y = 0$, $xz$-plane. This implies that while torsional motion exists on each side of the $y = 0$ plane, there is no net torsional motion around the spine. The integration area $-1 < x, y < 1$, results in a net circulation around the spine of zero. These results suggest that the torsional motion on each side manifested as vortex tubes adjacent to the spine, and are directly generated by the motion of the spine itself. This motion is mainly in the $x$ direction, anchored at the null point, and its amplitude decreases along the $z$ direction.

Figure \ref{fig:circulation-diagram}a presents a view of the vortices generated around the spine  at $t=$ 1 for baseline case $\psi=0.05$. The plot shows vortices identified by isosurfaces of the Q criterion \citep{Jeong_Hussain_1995} with the color scale indicating the size of $\omega_z = \pmb{\omega} \cdot \hat{\bf{z}}$. The Q criterion identify vortices as regions where the vorticity magnitude exceeds the magnitude of the strain rate tensor. We observe  that there are two counter-rotating vortices for $z>0$ and another two for $z<0$. The black line represents the spine and the vortex rolls on either side of the spine are generated by the spine movement.

Figure \ref{fig:circulation-diagram}b  displays contour plots of $\Gamma(z,t)$ in a time-distance diagram. Red crosses denote the roots of $j_y(0,0,0,t)$ (as seen in Figure \ref{fig:jy-fitting}). The time-distance diagram reveals that the oscillations in circulation are synchronized with the roots of $j_y(0,0,0,t)$, with the $\Gamma$ sign changing as the $j_y(0,0,0,t)$ sign changes. The circulation follows a vertical trajectory that reverses rotation across the fan plane. We also take the average of $v_s/v_a$ over the $xy$-plane at each position along the $z$ axis, with the black lines indicating where this value is equal to one (i.e. an averaged equipartition layer). 
Inside this averaged equipartition layer the circulation is slightly more intense than outside of it, and this stronger circulation can be linked to the extra small vortex rolls near the null point displayed in Figure \ref{fig:circulation-diagram}a. 

Figure \ref{fig:circulation-diagram}c displays the circulation sampled along the dashed lines in Figure \ref{fig:circulation-diagram}b. It demonstrates that the circulation signal at different heights shares the same amplitude and exhibits no phase shift, indicating the absence of $\omega_z$ propagating along the $z$-direction. The amplitude of $\Gamma$ is strongly damped after $t \approx 30$ as shown in Figure \ref{fig:circulation-diagram}b and \ref{fig:circulation-diagram}c. This damping is related to the damping of $j_y(0,0,0,t)$ in Figure \ref{fig:jy-fitting}.

%%%%%%%%%%%%%%%%%%%%%%%%%%%%%%%%%%%%%%%%%%%%%%

\section{Conclusions} \label{sec:conclusions}

This paper investigates the long-term periodic behavior of oscillatory reconnection generated from a three-dimensional, linear, proper, potential null point. The three-dimensional null point was perturbed by an initial condition to the magnetic field, which in turn triggered the oscillatory reconnection phenomena, i.e. a magnetic relaxation process. We investigated four different amplitudes, $\psi$, of our initial disturbance, with $\psi = 0.1$, the largest perturbation studied, corresponding to a significant bending of the spine and a twist around the null point, and with $\psi=0.01$, the smallest perturbation studies, corresponding to the least amount of bending of the spine.

Via tracking the location of the null point, it was confirmed that the null point remains stationary, consistently located at $(x, y, z) = (0, 0, 0)$ throughout the entire evolution. This behavior persisted across all perturbation amplitudes due to the symmetric nature of the applied perturbation.

We observe a clear signature of oscillatory reconnection, characterized by current density oscillations at the null point, $j_y(0,0,0,t)$, lasting up to 60 Alfvén time scales. We observe periodic behavior in $j_y(0,0,0,t)$, characterized by cycles of current sheets in what we call \lq{orientation one}\rq{} for $j_y(0,0,0,t)>0$ followed by current sheets in \lq{orientation two}\rq{} for $j_y(0,0,0,t)<0$. Multiple repetitions of this cyclic behavior are observed.

We investigated four different amplitudes for the initial perturbation ($\psi=0.01$, $\psi=0.025$, $\psi=0.05$ and $\psi=0.1$). %We find a 
An invariant solution exists for $j_y(0,0,0,t)$ when it is normalized to the initial perturbation amplitude $\psi$. The overall behavior of the system remains consistent across all four cases, but the amplitude varies with $\psi$.

The $j_y(0,0,0,t)/\psi$ displays an invariant behavior, with the same period for all simulation cases. Thus the oscillation period is independent of the initial pulse amplitude. A period of $8.1 \pm 1.1$ Alfv\'en times is extracted using a Morlet wavelet for $\psi=0.05$ and a time window of $t=20-60$. A near-identical period (all within errors bars) is extracted for different values of $\psi$ and for different time windows. In addition to this, a near-identical period (all within errors bars) is found when the period is extracted via Fourier spectra. Thus, we conclude this invariant system is characterized by a single periodicity of $8.1 \pm 1.1$ 
 Alfv\'en times. 

As detailed in Section 2.1, our simulation results can be scaled with appropriate reference scales, and for typical values for the solar corona of $L_0=1$ Mm, $B_0=1$G and $\rho_0 = 1.67 \times 10^{-12}$ kg/m$^{3}$ this gives $t_0=L_0 \sqrt{\mu_0 \rho_0}/B_0 =  14.4865$ s. This means our system would have a period of $(8.1\pm 1.1)t_0=117.3\pm15.9$ s. However, our choice of equilibrium magnetic field is scale-free. This freedom in setting $B_0$ and $L_0$, and hence the choice of $t_0$, is not unique. For this reason we caution reading too much into this dimensional period. 
Note that in our system the magnetic Reynolds number is non-dimensionalised such that $\eta_0= \mu_0L_0v_0$ (section \ref{sec2.1}) and thus our system can be rescaled if one keeps $\eta_0$ the same. However \cite{Talbot2024} found for a 2D null point that the period is independent of the resistivity and thus if this result transfers over to 3D then our simulation results can be rescaled independent of this constrain on $\eta_0$.

The behavior of the normalized current density $j_y(0,0,0,t)/\psi$ is invariant, which implies that the damping rate is independent of the initial perturbation amplitude. 
$j_y(0,0,0,t)$ was fitted with a periodic decaying signal using six different possible functions: either a sine wave or a Bessel function to capture the oscillation, and either an exponential decay, a Gaussian decay or a Generalized Damping Profile (GDP) to capture the decay. It was found that all options give a good fit to  $j_y(0,0,0,t)$ (i.e. lowest root-mean-square in the root positions of $j_y(0,0,0,t)$, and the highest coefficient-of-determination $R^2$), with the optimum fit 
provided by the sine-wave oscillation with the Gaussian envelope, i.e. $ a \sin\left[ {2\pi} (t-\phi) / \Lambda \right]  \exp\left(-{t^2} / {2 \tau_g^2}\right)$.

%We also report 
It was found that the $j_y(0,0,0,t)$ decays back to zero after $t=60$ Alfv\'en times, i.e. returns to initial equilibrium. This is in contrast to the two-dimensional case \citep{McLaughlin2009}, where the asymmetric plasma heating drives the system toward a non-zero $| \bf{j} |$ value at the null. This is because, in 2D, the separatrices do not allow the %generated reconnection-heat 
heat generated by the reconnection 
to spread evenly around the null%: i.e. 
; the separatrices divide the 2D region into four domains of connectivity, and heat is not allowed to conduct across these different connectivities. Therefore pressure gradients are allowed to build-up, leading to the 2D null being slightly \lq{scissored-up}\rq{} at the end of the simulation and thus slightly non-potential, i.e. a non-zero $| \bf{j} |$ value at the null. However, in 3D,  the heating spreads more evenly along and around the fan plane. This manifest as negligible pressure gradient across the fan plane, thus creating a negligible $j_y(0,0,0,t=60)$ at the end of the simulation.

This is the longest duration signal generated by oscillatory reconnection reported around a three-dimensional null point to date.  \cite{Thurgood2017} were the first to investigate oscillatory reconnection in 3D using the $\psi=0.05$ simulation reported here. However, their simulations were limited to 6 Alfvén time scales, i.e. only one oscillation period, as this was the point at which reflected waves from the boundary reached their null point. Their study did not incorporate a damping mechanism to mitigate the effects of reflected waves on the oscillatory reconnection process, which constrained the duration of their simulations. In contrast, our implementation of a damping procedure (the sponge boundaries) allowed us to extend the simulation time significantly, providing new insights into the long-term behavior of oscillatory reconnection in 3D. In addition, we investigate four different amplitudes for the initial perturbation ($\psi=0.01$, $\psi=0.025$, $\psi=0.05$ and $\psi=0.1$) whereas \cite{Thurgood2017} only considered $\psi=0.05$.

Note that $j_y(0,0,0,t)/\psi$ curves in the early phases are offset in time. This is because the strongest perturbation initiates the reconnection cycle earlier than the cases with smaller $\psi$. After this, 
the system’s invariance, i.e. $\psi$ affecting the amplitude of $j_y(0,0,0,t)$ but not the period, dominates. This short-lived transient followed by an invariant signal hints at a single  characteristic period being associated with this three-dimensional null point, inviting the possibility of using the period as a diagnostic of the underlying null point properties. This is similar to that observed in two-dimensional configurations by \cite{Karampelas2022b}.

Finally, to assess potential torsional dynamics, we computed the circulation as a proxy for net azimuthal motion around the spine at multiple heights. Our analysis revealed no net torsional motion, indicating that localized vortex tubes adjacent to the spine arise solely from the motion of the spine, rather than from large-scale twisting.

There are many directions this work can be taken in the future. For example, in future work we will investigate characteristics of waves generated by the reconnection, including their dependence on magnetic field and plasma parameters, utilizing modal decomposition techniques. In addition, we will investigate the influence of thermal conduction, building upon the 2D null point results of \cite{Karampelas2022a,Karampelas2023}.

%%%%%%%%%%%%%%%%%%%%%%%%%%%%%%%%%%%%%%%%%%%%%%%%%%%%%%%%%%%%%%%%%%%%%%%%%%%%%%%%
%%%%%%%%%%%%%%%%%%%%%%%%%%%%%%%%%%%%%%%%%%%%%%%%%%%%%%%%%%%%%%%%%%%%%%%%%%%%%%%%

%%%%%%%%%%%%%%%%%%%%%%%%%%%%%%%%%%%%%%%%%%%%%%%%%%%%%

%\begin{acknowledgments}
\section*{Acknowledgments}
All authors acknowledge the UK Research and Innovation (UKRI) Science and Technology Facilities Council (STFC) for support from grant ST/X001008/1 and for IDL support. The research was sponsored by the DynaSun project and has thus received funding under the Horizon Europe programme of the European Union under grant agreement (no. 101131534). Views and opinions expressed are however those of the author(s) only and do not necessarily reflect those of the European Union and therefore the European Union cannot be held responsible for them. This work was also supported by the Engineering and Physical Sciences Research Council (EP/Y037464/1) under the Horizon Europe Guarantee. This work used the Oswald High Performance Computing facility operated by Northumbria University (UK), and the DiRAC Data Intensive service (CSD3) at the University of Cambridge, managed by the University of Cambridge University Information Services on behalf of the STFC DiRAC HPC Facility (www.dirac.ac.uk). The DiRAC component of CSD3 at Cambridge was funded by BEIS, UKRI and STFC capital funding and STFC operations grants. DiRAC is part of the UKRI Digital Research Infrastructure. Numerical simulations were conducted with LARE3D which is available at \href{https://github.com/Warwick-Plasma/Lare3d}{https://github.com/Warwick-Plasma/Lare3d}. The data that support the findings of this study are available from the corresponding author upon reasonable request.
%\end{acknowledgments}

\software{LARE3D \citep{Arber2001},
          NumPy \citep{numpy},  
          SciPy \citep{SciPy}, 
          Matplotlib \citep{matplotlib}
          WaLSA tools \citep{jafarzadeh_wave_2025}.
          }

%%%%%%%%%%%%%%%%%%%%%%%%%%%%%%%%%%%%%%%%%%%%%%%%%%%%
% \appendix

\bibliography{references}{}
\bibliographystyle{aasjournal}

%% This command is needed to show the entire author+affiliation list when
%% the collaboration and author truncation commands are used.  It has to
%% go at the end of the manuscript.
%\allauthors

%% Include this line if you are using the \added, \replaced, \deleted
%% commands to see a summary list of all changes at the end of the article.
%\listofchanges

\end{document}